\documentclass[amsmath,amssymb,prd,nofootinbib,superscriptaddress,12pt]{revtex4-2}

\usepackage{amsfonts}
\usepackage{bm}
\usepackage{mathrsfs}
\usepackage{siunitx}
\usepackage{booktabs}
\usepackage{array}
\usepackage{multirow}
\usepackage{makecell}
\usepackage{tabularx}
\usepackage{adjustbox}

\usepackage{graphicx}
\usepackage{xcolor}
\usepackage{float}
\usepackage{rotating}
\usepackage{caption}
\usepackage{subcaption}
\usepackage[a4paper,margin=1.6cm]{geometry}

\usepackage{tensor}
\usepackage{bigints}
\usepackage{bbm}
\usepackage{hyperref}
\hypersetup{
  colorlinks=true,
  breaklinks=true,
  citecolor=blue,
  linkcolor=[rgb]{0,0.5,0.9},
  urlcolor=red,
  filecolor=green
}

\begin{document}

\title{\Large{Periodic orbits and gravitational wave radiation of a phantom Dirac--Born--Infeld regular black hole Gravity}}

\author{N.~Heidari
}
\email{heidari.n@gmail.com (The corresponding author)}

\affiliation{Departamento de Física, Universidade Federal de Campina Grande Caixa Postal 10071, 58429-900 Campina Grande, Paraíba, Brazil.}

\affiliation{Center for Theoretical Physics, Khazar University, 41 Mehseti Street, Baku, AZ-1096, Azerbaijan.}



\author{A.~A.~Ara\'{u}jo~Filho
}
\email{dilto@fisica.ufc.br}

\affiliation{Departamento de Física, Universidade Federal de Campina Grande Caixa Postal 10071, 58429-900 Campina Grande, Paraíba, Brazil.}
\affiliation{Center for Theoretical Physics, Khazar University, 41 Mehseti Street, Baku, AZ-1096, Azerbaijan.}
\affiliation{Departamento de Física, Universidade Federal da Paraíba, Caixa Postal 5008, 58051--970, João Pessoa, Paraíba,  Brazil.}

\author{Iarley P. Lobo
}
\email{lobofisica@gmail.com}

\affiliation{Departamento de Física, Universidade Federal da Paraíba, Caixa Postal 5008, 58051--970, João Pessoa, Paraíba,  Brazil.}
\affiliation{Departamento de Física, Universidade Federal de Campina Grande Caixa Postal 10071, 58429-900 Campina Grande, Paraíba, Brazil.}

\author{V. B. Bezerra
}
\email{valdir@fisica.ufpb.br}
\affiliation{Departamento de Física, Universidade Federal da Paraíba, Caixa Postal 5008, 58051--970, João Pessoa, Paraíba,  Brazil.}

\begin{abstract}

We investigate periodic orbits and gravitational wave radiation in the spacetime of an asymptotically flat regular black hole supported by a phantom Dirac--Born--Infeld scalar field. Starting from the timelike geodesic equations, we examine the effective potential and the characteristic circular orbits and construct periodic zoom--whirl trajectories. We further explore the sensitivity of orbital closure to small energy perturbations and examine the periastron advance of the S2 star around Sgr~A$^*$ to assess possible constraints on the regular core parameter. The gravitational wave signals associated with representative periodic trajectories are then calculated within a quadrupole-based numerical-kludge approximation, considering both polarization modes. Particular attention is given to the influence of the regular core on orbital configurations and waveform characteristics, providing a complementary perspective on gravitational aspects of phantom Dirac--Born--Infeld regular black holes.

\end{abstract}
\maketitle

\pagebreak

\tableofcontents


\section{Introduction}
\label{sec:introduction}

The possibility of resolving black hole singularities without eliminating their characteristic exterior geometry has motivated considerable interest in regular black hole solutions. Although the classical singularity theorems predict geodesic incompleteness under specific assumptions, the ultimate behavior of gravitational collapse remains uncertain when these conditions are relaxed. Regular black holes provide an alternative description in which the singular region is replaced by a smooth core, while an event horizon may still be present. Beyond their theoretical significance, these geometries offer a means of investigating how modifications of the central region influence particle dynamics and gravitational radiation.

A recent example was proposed by Parvez and Shankaranarayanan \cite{parvez2026exact}, who obtained an exact, asymptotically flat solution in Einstein gravity coupled to a nonlinear Dirac--Born--Infeld (DBI) scalar field. By adopting a regularity motivated areal radius, the gravitational field equations lead to a nonsingular geometry supported by the phantom branch of the scalar theory. The resulting spacetime possesses scalar hair, approaches the Schwarzschild solution at large distances, and is characterized by a regular core parameter. The original construction also suggested a possible connection between its limiting configurations, primordial black hole remnants, and dark matter.

Some properties of this newly proposed spacetime have been explored recently. Its thermodynamics, including Hawking temperature, heat capacity, radiation sparsity, and energy emission, were investigated in Ref.~\cite{ahmed2026hawking}. Quasinormal modes of massive scalar perturbations and massless fields of different spins were examined in Refs.~\cite{skvortsova2026massive,lutfuouglu2026scalar}. Eikonal ringing, black hole shadows, gravitational lensing, greybody factors, and circular-orbit energetics were studied in Ref.~\cite{abdullaev2026eikonal}, while particle creation, energy-conserving emission, wave absorption and scattering, and geodesic motion were addressed in Ref.~\cite{heidari2026exploring}. Despite these developments, the periodic orbital structure of this geometry and its associated gravitational wave emission have not yet been systematically investigated.

Bound timelike geodesics offer a direct probe of the strong--field gravitational regime. Among them, periodic orbits are particularly informative because their radial and azimuthal motions become commensurate, producing closed trajectories after a finite number of radial oscillations. Near the boundary of orbital instability, these trajectories may develop characteristic zoom--whirl patterns, in which the particle performs several nearly circular revolutions before returning toward apastron. Such behavior is sensitive to the underlying effective potential and can reveal subtle departures from Schwarzschild dynamics. Periodic geodesics also provide useful reference configurations for studying gravitational radiation from extreme mass-ratio systems, where the orbital motion can be approximated as geodesic over timescales short compared with radiation-reaction effects. The resulting waveforms establish a connection between strong--field orbital dynamics and the geometry of the central compact object \cite{Addazi:2021xuf}.

Periodic trajectories and their gravitational wave signatures have been investigated in various black hole backgrounds, including static and rotating regular black holes \cite{Gong:2025mne,Kumar:2024our}, noncommutative geometries \cite{HEIDARI2026170773,Ahmed:2025azu}, quantum-corrected spacetimes \cite{Chen:2025aqh,Ahmed:2025shr}, and effective field theory models \cite{Alloqulov:2025dqi}. Related studies have considered the $\gamma$ metric \cite{Zhang:2025wni}, Einstein--Aether and bumblebee gravity \cite{Lu:2025cxx,Shi:2026zxx,Liu:2025swi}, Kalb--Ramond black holes \cite{Junior:2024tmi,Xia:2025yzg}, and black holes surrounded by dark matter \cite{heidari2026gravitational}. Further examples include charged black holes with scalar hair \cite{Deng:2025wzz}, renormalization-group-improved rotating geometries \cite{Li:2025sfe,kumar2024exploring}, black holes without a Cauchy horizon \cite{Wang:2025hla}, and other modified black hole spacetimes \cite{Hua:2026kvw,Xamidov:2026kqs}. These investigations demonstrate the sensitivity of orbital configurations and gravitational wave signals to modifications of the background geometry.

In this work, we investigate periodic timelike orbits and the corresponding gravitational wave emission around a regular black hole supported by a phantom DBI scalar field. We first analyze the effective potential and determine the marginally bound orbit (MBO) and innermost stable circular orbit (ISCO), which characterize the boundaries of the relevant orbital regime. We then construct periodic zoom--whirl trajectories and examine how their energies, angular momenta, and spatial configurations respond to variations in the regular core parameter. Small deviations from the energies of exactly periodic orbits are also considered to explore the resulting changes in orbital closure.

Finally, we study gravitational radiation from a compact body orbiting a much more massive black hole, calculating both gravitational wave polarizations within the quadrupole approximation. By comparing orbits of the same classification for different values of $a$, we investigate changes in waveform amplitude, temporal behavior, and periodicity relative to the Schwarzschild limit. In this way, our analysis extends previous investigations of phantom DBI regular black holes to periodic strong--field motion and its gravitational wave signatures, providing further insight into how the regular core affects the dynamical properties of the spacetime.


\section{Background geometry}
\label{sec:geometry}

We consider the asymptotically flat regular black hole solution introduced in Ref.~\cite{parvez2026exact}, arising from four dimensional Einstein gravity minimally coupled to a nonlinear Dirac--Born--Infeld (DBI) scalar field. The underlying theory is described by the action
\begin{equation}
S=\int \mathrm{d}^{4}x\sqrt{-g}
\left[
\frac{\mathcal{R}}{2\kappa^{2}}
+\epsilon V(\phi)
\left(
1-\sqrt{1+\frac{2X}{\Lambda^{4}}}
\right)
\right],
\label{eq:DBI_action}
\end{equation}
where
\begin{equation}
X=-\frac{1}{2}g^{\mu\nu}
\nabla_{\mu}\phi\nabla_{\nu}\phi,
\qquad
\kappa^{2}=8\pi G.
\end{equation}
Here $\mathcal{R}$ denotes the Ricci scalar, $\Lambda$ characterizes the DBI energy scale, and $V(\phi)$ determines the scalar-field coupling within the nonlinear sector. The parameter $\epsilon=\pm1$ specifies the sign of the scalar contribution. In the weak-gradient regime, $|X|/\Lambda^{4}\ll1$, the scalar Lagrangian reduces to
\begin{equation}
\mathcal{L}_{\phi}
\simeq
-\epsilon\frac{V(\phi)}{\Lambda^{4}}X
+\mathcal{O}\left(\frac{X^{2}}{\Lambda^{4}}\right).
\label{eq:DBI_low_energy}
\end{equation}
For positive $V(\phi)$, the choice $\epsilon=-1$ gives the conventional kinetic sign, whereas $\epsilon=+1$ corresponds to the phantom branch. The latter is required to support the regular geometry considered here \cite{parvez2026exact}.

The spacetime is static and spherically symmetric, with line element
\begin{equation}
\mathrm{d}s^{2}
=-f(r)\mathrm{d}t^{2}
+\frac{\mathrm{d}r^{2}}{f(r)}
+R^{2}(r)
\left(
\mathrm{d}\theta^{2}
+\sin^{2}\theta\,\mathrm{d}\varphi^{2}
\right),
\label{eq:DBImetric}
\end{equation}
where 
\begin{equation}
R(r)=\sqrt{r^{2}+a^{2}},
\label{eq:areal_radius}
\end{equation}

where the parameter \(a\), with dimensions of length, characterizes the regular core of the spacetime and sets the minimum areal radius, \(R(0)=a\). The corresponding two-sphere therefore has a finite area, $4\pi a^{2}$. This distinction between the radial coordinate and the areal radius must be retained when describing particle motion and geometrical quantities.

For the choice in Eq.~\eqref{eq:areal_radius}, the combination of Einstein equations obtained from $G^{t}_{t}=G^{\theta}_{\theta}$ is independent of the scalar profile and reduces to
\begin{equation}
\left(r^{2}+a^{2}\right)f''(r)
-2f(r)+2=0.
\label{eq:lapse}
\end{equation}
Imposing asymptotic flatness and fixing the coefficient of the inverse-radius term by the ADM mass $M$ yields \cite{parvez2026exact}
\begin{equation}
f(r)=1+\frac{3GM}{a}
\left[
\frac{r}{a}
-\left(1+\frac{r^{2}}{a^{2}}\right)
\tan^{-1}\left(\frac{a}{r}\right)
\right].
\label{eq:exactlapse}
\end{equation}
The remaining field equations determine the scalar configuration and select the phantom branch as the consistent matter source. Thus, the metric is obtained by solving the gravitational equations after specifying the regular areal-radius profile, rather than by introducing an arbitrary deformation of the Schwarzschild lapse.

The horizon structure is controlled by the ratio between the mass and the regular core parameter. The event horizon is located at the positive root of $f(r_h)=0$, while the limiting value of the lapse at the core is
\begin{equation}
f(0)=1-\frac{3\pi GM}{2a}.
\label{eq:core_lapse}
\end{equation}
The nonextremal black hole branch exists for
\begin{equation}
M>\frac{2a}{3\pi G}.
\label{eq:DBI_BH_condition}
\end{equation}
At equality, the horizon approaches the minimal-radius surface, whereas the continuation beyond this boundary does not belong to the nonextremal black hole branch. Our analysis is restricted to configurations possessing an event horizon.

Henceforth, we adopt geometrized units $G=c=\hbar=1$. At distances large compared with $a$, the metric functions admit the expansions
\begin{equation}
R(r)=r+\frac{a^{2}}{2r}
+\mathcal{O}(r^{-3}),
\label{eq:R_asymptotic}
\end{equation}
and
\begin{equation}
f(r)=1-\frac{2M}{r}
+\frac{2Ma^{2}}{5r^{3}}
+\mathcal{O}(r^{-5}).
\label{eq:fasymptotic}
\end{equation}
The leading asymptotic term therefore reproduces the Schwarzschild mass contribution, while the first correction to the lapse is quadratic in $a$ and enters at order $r^{-3}$.


\section{Timelike geodesic motion }

To explore the orbital properties of the spacetime, we consider the motion of a neutral massive test particle following a timelike geodesic. Since the background is static and spherically symmetric, the dynamics can be conveniently described within the Hamiltonian formalism, with the conserved quantities.
The Hamiltonian of a particle with rest mass \(m\) is written as
\begin{equation}
\mathcal{H}
=
\frac{1}{2}
\left(
g^{\mu\nu}p_{\mu}p_{\nu}
+
m^{2}
\right),
\label{eq:hamiltonian}
\end{equation}
where \(p_{\mu}\) denotes the canonical four-momentum. 
For the static and spherically symmetric geometry considered here, the Hamiltonian takes the explicit form
\begin{equation}
\mathcal{H}
=
\frac{1}{2}
\left[
-\frac{p_{t}^{2}}{f(r)}
+
f(r)p_{r}^{2}
+
\frac{p_{\theta}^{2}}{R^{2}(r)}
+
\frac{p_{\phi}^{2}}{R^{2}(r)\sin^{2}\theta}
+
m^{2}
\right].
\label{eq:explicit_hamiltonian}
\end{equation}

The equations of motion follow from Hamilton's equations,
\begin{equation}
\frac{dx^{\mu}}{d\lambda}
=\frac{\partial \mathcal{H}}{\partial p_{\mu}},
\qquad
\frac{dp_{\mu}}{d\lambda}
=-\frac{\partial \mathcal{H}}{\partial x^{\mu}},
\label{eq:hamilton_equations}
\end{equation}
where \(\lambda\) is an affine parameter. 

For a massive particle, it is convenient to use the proper time \(\tau\) with $p^{\mu}=
m \frac{dx^{\mu}}{d\tau}$, the mass-shell condition is $g^{\mu\nu}p_{\mu}p_{\nu}
=
-m^{2}$ and therefore the physical trajectory satisfies $\mathcal{H}=0$. 

We introduce the specific energy \(E\) and the angular momentum \(L\) according to
\begin{equation}
E
=-\frac{p_{t}}{m},
\qquad
L
=
\frac{p_{\phi}}{m}.
\label{eq:energy_angular_momentum}
\end{equation}
Using these definitions, the components of the four-velocity become
\begin{equation}
\frac{dt}{d\tau}=
\frac{E}{f(r)},
\label{eq:t_equation}
\end{equation}

\begin{equation}
\frac{d\phi}{d\tau}=\frac{L}{R^{2}(r)\sin^{2}\theta},
\label{eq:phi_equation}
\end{equation}

\begin{equation}
\frac{d\theta}{d\tau}=\frac{p_{\theta}}{m R^{2}(r)},
\label{eq:theta_equation}
\end{equation}
and the radial equation can be written as
\begin{equation}
\frac{dr}{d\tau}
=\sqrt{E^{2}-
f(r)
\left[
1+
\frac{L^{2}}{R^{2}(r)\sin^{2}\theta}
+
\frac{p_{\theta}^{2}}{m^{2}R^{2}(r)}
\right].}
\label{eq:radial_equation}
\end{equation}

Considering the geodesic motion restricted within the equatorial plane, $\theta=\frac{\pi}{2}$, and $
p_{\theta}=0$, the radial equation then reduces to
\begin{equation}
\left(
\frac{dr}{d\tau}
\right)^{2}
+
V_{\rm eff}=E^{2},
\label{eq:Veff}
\end{equation}
where the effective potential is given by
\begin{equation}
V_{\rm eff}=f(r)
\left(
1
+
\frac{L^{2}}{r^{2}+a^{2}}
\right).
\label{eq:Veff}
\end{equation}

The parameter \(a\) therefore modifies the particle dynamics through the lapse function \(f(r)\) and the nontrivial areal radius \(R(r)=\sqrt{r^{2}+a^{2}}\). This feature distinguishes the present geometry from the Schwarzschild spacetime, for which \(R(r)=r\). In the limit \(a\rightarrow 0\), the standard Schwarzschild effective potential is recovered.



\begin{figure}[ht!]
    \centering
    \includegraphics[width=80mm]{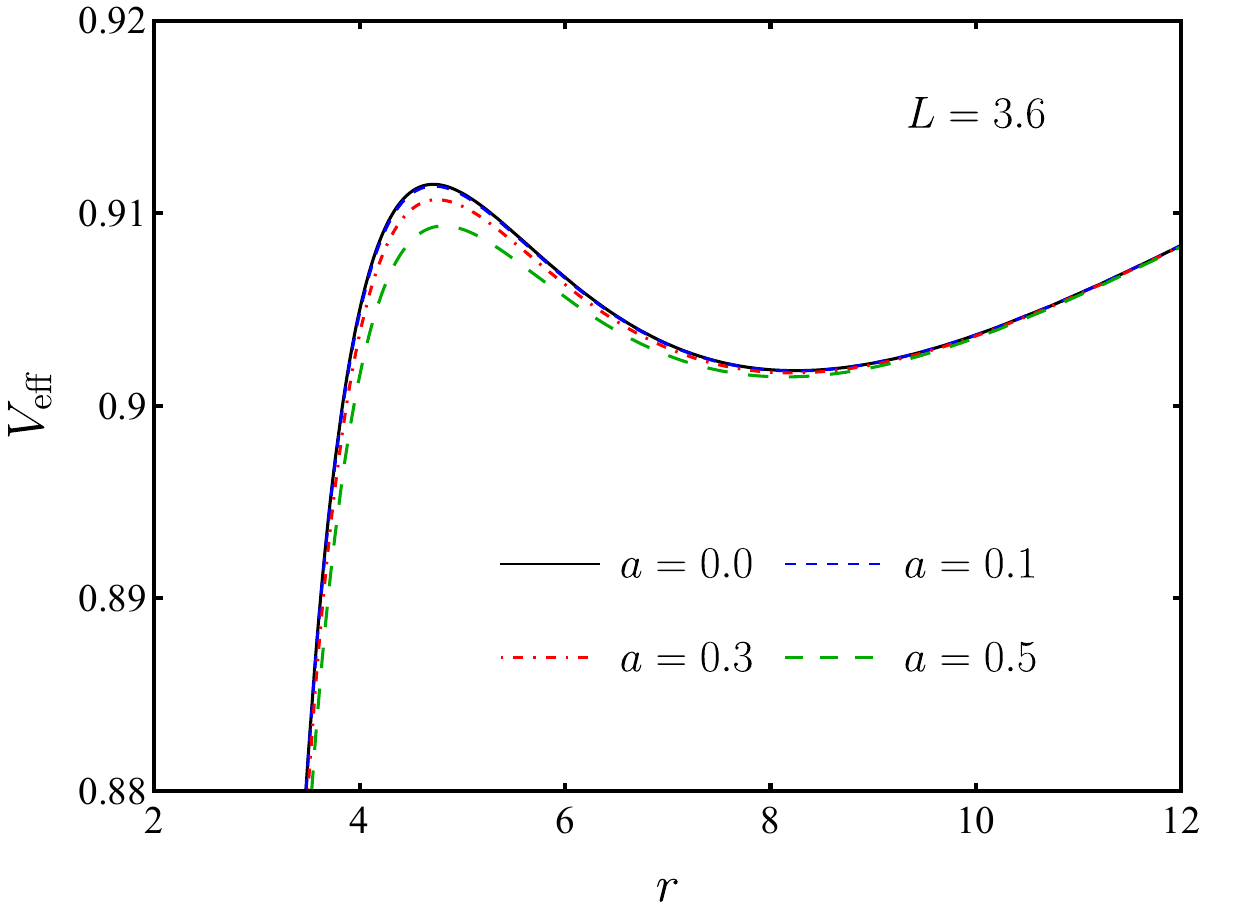}
    \includegraphics[width=80mm]{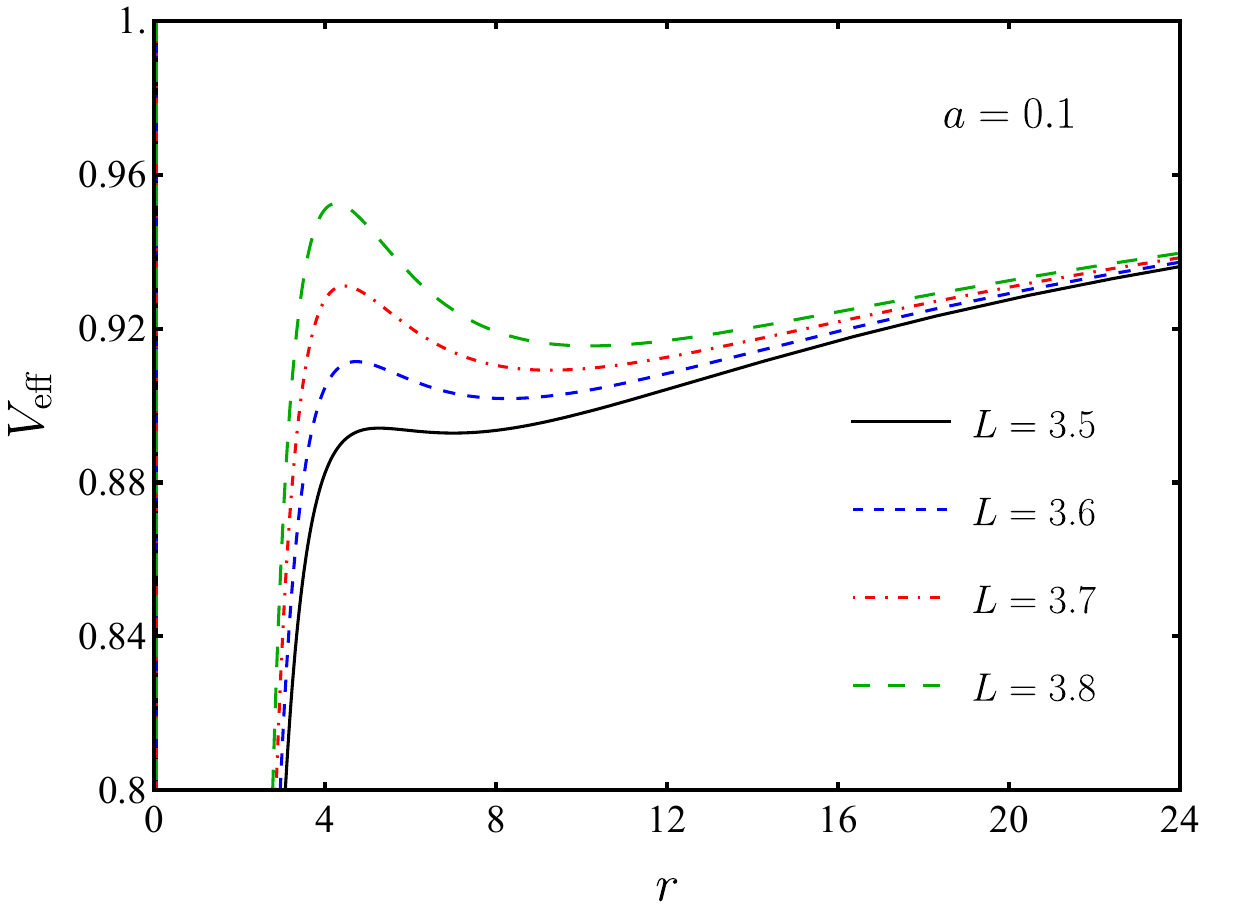}
    \protect\caption{Dependence of the radial behavior of the effective potential on the parameter $a$ and the angular momentum $L$, with $M=1$. The left panel shows the variation with $a$ for fixed $L=3.6$, while the right panel illustrates the dependence on $L$ for fixed $a=0.1$.
}
    \label{fig:Veff}
\end{figure}

The behavior of the effective potential is illustrated in Fig.~\ref{fig:Veff}. 
For fixed angular momentum, $L=0.36$, increasing the parameter $a$ produces only a mild modification of the potential, mainly around its local maximum. In particular, the height of the potential barrier increases slightly with $a$, while the curves approach each other at large distances. This indicates that the effect of $a$ is concentrated in the strong--field region close to the black hole. The dependence on the angular momentum is considerably stronger. For fixed $a=0.1$, increasing $L$ raises the effective potential and makes the barrier more pronounced. At the same time, the local maximum moves toward smaller radii, whereas the minimum associated with stable circular motion shifts outward. Thus, the angular momentum plays a more significant role in shaping the structure of the effective potential, while the parameter $a$ introduces a comparatively small correction over the range considered.
 
The asymptotic behavior of the effective potential provides a useful characterization of the allowed timelike trajectories. Since \(V_{\rm eff}\rightarrow 1\) as \(r\rightarrow\infty\), the value \(E=1\) separates bound from unbound motion: particles with \(E>1\) may reach spatial infinity, whereas bound trajectories require \(E<1\). Stable bound motion is therefore restricted by the characteristic energies and angular momenta associated with the innermost stable circular orbit (ISCO) and the marginally bound orbit (MBO), so that
\begin{equation}
E_{\rm ISCO}\leq E\leq E_{\rm MBO}=1,
\qquad
L\geq L_{\rm ISCO}.
\label{eq:boundcond}
\end{equation}
Here, the MBO represents the limiting unstable circular orbit for which the particle has just enough specific energy to escape to infinity, while the ISCO marks the inner boundary of stable circular motion. These two limiting orbits therefore play a central role in delimiting the region in which bound trajectories can exist. In what follows, we are particularly interested in periodic bound orbits, for which the radial and azimuthal motions satisfy a commensurability condition, causing the trajectory to close after a finite number of orbital cycles. Such orbits provide a useful probe of the underlying spacetime geometry, since changes in the regular core parameters modify the effective potential and, consequently, the location and structure of the allowed orbital configurations.

Before discussing periodic trajectories, it is useful to identify the
characteristic circular orbits that set the boundaries of bound timelike
motion. In particular, the marginally bound orbit (MBO) and the innermost
stable circular orbit (ISCO) provide natural reference scales for the orbital
dynamics. Since the spacetime is asymptotically flat, the effective potential
approaches unity at spatial infinity,
\begin{equation}
\lim_{r\rightarrow\infty} V_{\mathrm{eff}}(r)=1.
\label{eq:Veff_infinity}
\end{equation}
Hence, $E=1$ marks the threshold between bound and unbound motion. The MBO
corresponds to this limiting energy.
\begin{equation}
V_{\mathrm{eff}}(r_{\mathrm{MBO}})=1,
\qquad
\left.
\frac{dV_{\mathrm{eff}}}{dr}
\right|_{r=r_{\mathrm{MBO}}}
=0.
\label{eq:MBO_conditions}
\end{equation}
Substituting Eq.~\eqref{eq:Veff} into these conditions gives
\begin{equation}
\left(r_{\mathrm{MBO}}^{2}+a^{2}\right)
f'(r_{\mathrm{MBO}})
-
2r_{\mathrm{MBO}}f(r_{\mathrm{MBO}})
\left[1-f(r_{\mathrm{MBO}})\right]
=0.
\label{eq:MBO_radius}
\end{equation}
Once $r_{\mathrm{MBO}}$ is obtained, the corresponding specific angular
momentum follows from
\begin{equation}
L_{\mathrm{MBO}}
=
\sqrt{
\left(r_{\mathrm{MBO}}^{2}+a^{2}\right)
\frac{1-f(r_{\mathrm{MBO}})}
{f(r_{\mathrm{MBO}})}
}.
\label{eq:MBO_angular_momentum}
\end{equation}

On the other hand, the ISCO corresponds
to a circular orbit that is marginally stable against radial perturbations.
Its location is therefore obtained by imposing
\begin{equation}
V_{\mathrm{eff}}(r_{\mathrm{ISCO}})
=
E_{\mathrm{ISCO}}^{2},
\quad
\left.
\frac{dV_{\mathrm{eff}}}{dr}
\right|_{r=r_{\mathrm{ISCO}}}
=0,
\quad
\left.
\frac{d^{2}V_{\mathrm{eff}}}{dr^{2}}
\right|_{r=r_{\mathrm{ISCO}}}
=0.
\label{eq:ISCO}
\end{equation}
Taking into account the nontrivial areal radius in Eq. \eqref{eq:areal_radius}, the marginal stability condition becomes
\begin{align}
&r_{\mathrm{ISCO}}
\left(r_{\mathrm{ISCO}}^{2}+a^{2}\right)
f(r_{\mathrm{ISCO}})
f''(r_{\mathrm{ISCO}})
-
2r_{\mathrm{ISCO}}
\left(r_{\mathrm{ISCO}}^{2}+a^{2}\right)
\left(f'(r_{\mathrm{ISCO}})\right)^{2}\\
&
+
\left(3r_{\mathrm{ISCO}}^{2}-a^{2}\right)
f(r_{\mathrm{ISCO}})
f'(r_{\mathrm{ISCO}})
=0.
\label{eq:ISCO_radius}
\end{align}
The corresponding specific energy is
\begin{equation}
E_{\mathrm{ISCO}}
=
\sqrt{
\frac{
2r_{\mathrm{ISCO}}
\left(f(r_{\mathrm{ISCO}})\right)^{2}
}{
2r_{\mathrm{ISCO}}f(r_{\mathrm{ISCO}})
-
\left(r_{\mathrm{ISCO}}^{2}+a^{2}\right)
f'(r_{\mathrm{ISCO}})
}
}.
\label{eq:ISCO_energy}
\end{equation}
Likewise, the specific angular momentum is
\begin{equation}
L_{\mathrm{ISCO}}
=
\sqrt{
\frac{
\left(r_{\mathrm{ISCO}}^{2}+a^{2}\right)^{2}
f'(r_{\mathrm{ISCO}})
}{
2r_{\mathrm{ISCO}}f(r_{\mathrm{ISCO}})
-
\left(r_{\mathrm{ISCO}}^{2}+a^{2}\right)
f'(r_{\mathrm{ISCO}})
}
}.
\label{eq:ISCO_angular_momentum}
\end{equation}

Table~\ref{tab:MBO_ISCO} summarizes the characteristic quantities of the MBO and ISCO for different values of the regular core parameter, with $M=1$. The Schwarzschild results are recovered at $a=0$. As $a$ increases, both characteristic radii move slightly outward, accompanied by small increases in their angular momenta and the ISCO energy. Although these deviations remain modest over the parameter range considered, they indicate a systematic modification of the circular-orbit structure induced by the regular core.

\begin{table}[t]
\centering
\caption{Marginally bound orbit (MBO) and innermost stable circular orbit (ISCO) quantities for several values of the parameter $a$, assuming $M=1$. The Schwarzschild limit corresponds to $a=0$.}
\label{tab:MBO_ISCO}
\setlength{\tabcolsep}{10pt}
\begin{tabular}{cccccc}
\hline
$a$ & $r_{\mathrm{MBO}}$ & $L_{\mathrm{MBO}}$ & $r_{\mathrm{ISCO}}$ & $E_{\mathrm{ISCO}}$ & $L_{\mathrm{ISCO}}$ \\
\hline\hline
0 & 4.00000 & 4.00000 & 6.00000 & 0.942809 & 3.46410 \\
0.1 & 4.00100 & 4.00075 & 6.00233 & 0.942835 & 3.46477 \\
0.2 & 4.00399 & 4.00299 & 6.00931 & 0.942913 & 3.46679 \\
0.3 & 4.00893 & 4.00671 & 6.02089 & 0.943043 & 3.47013 \\
0.4 & 4.01578 & 4.01189 & 6.03699 & 0.943222 & 3.47478 \\
\hline
\end{tabular}
\end{table}


\begin{figure}[ht!]
    \centering
    \includegraphics[width=80mm]{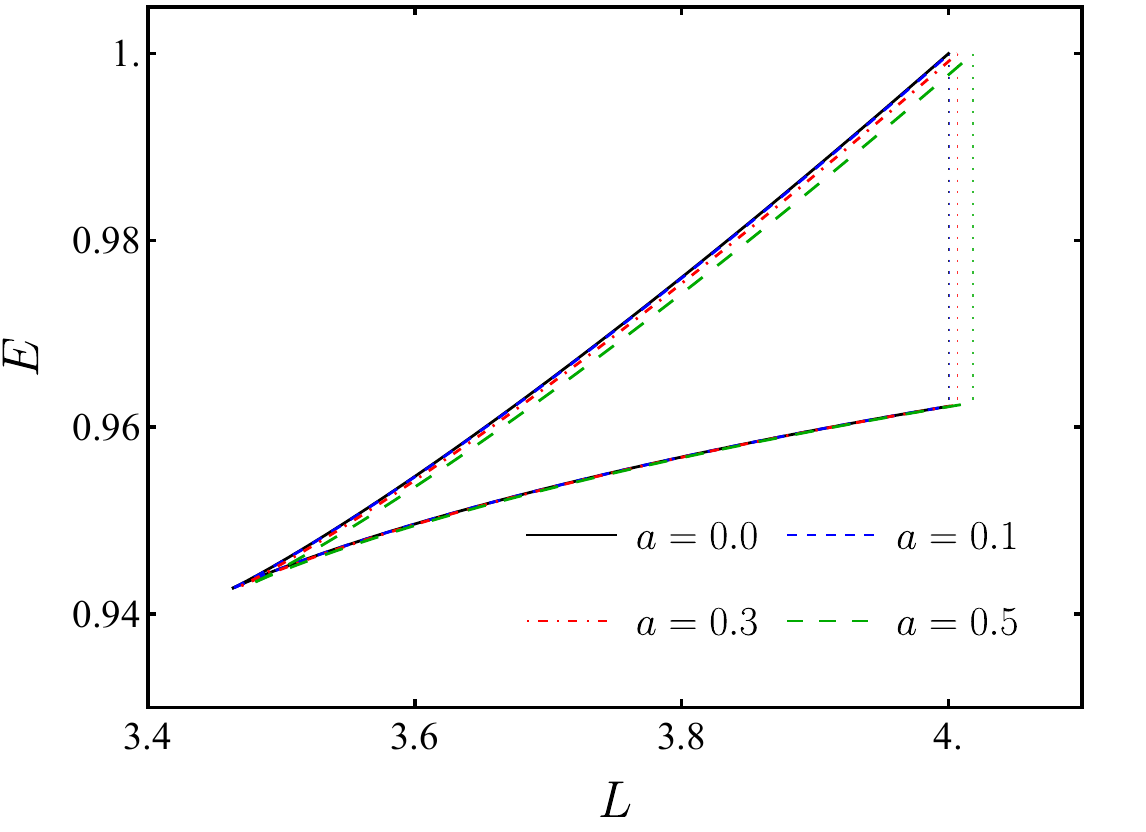} 
    \caption{For each value of $a$, the admissible region in the $(E, L)$ plane is the domain bounded by the corresponding colored curves, demonstrating the impact of altering $a$ while maintaining a constant mass of $M = 1$.}
    \label{fig:EL}
\end{figure}

To visualize the parameter range compatible with bound motion around the phantom DBI regular black hole, Fig.~\ref{fig:EL} displays the allowed region in the $(E,L)$ plane obtained from the conditions in Eq.~\eqref{eq:boundcond}. The results are shown for several values of the regular core parameter, $0\leq a\leq0.5$, with $M=1$. As $a$ increases, the allowed region shifts toward larger orbital angular momenta, while the upper energy bound remains fixed at $E_{\mathrm{MBO}}=1$. This indicates that a stronger regular core modification requires a larger angular momentum to sustain bound periodic motion. Conversely, for a fixed value of $L$, increasing the core parameter lowers the energy required for the particle to remain on a periodic orbit, for a fixed value of angular momentum.


\section{Periodic orbits}
\label{sec:periodic}

We next investigate the periodic motion of massive test particles around the phantom Dirac--Born--Infeld regular black hole. Periodic geodesics provide a useful framework for describing bound motion in strong gravitational fields, where relativistic precession can produce intricate zoom--whirl trajectories. A bound orbit closes when its radial and azimuthal frequencies are commensurate, whereas an irrational frequency ratio gives rise to a nonclosing, precessing trajectory. Since generic bound orbits can be approximated arbitrarily closely by periodic ones, this classification also provides a useful basis for studying orbital dynamics and the corresponding gravitational wave signatures.

Restricting the motion to the equatorial plane, we denote the radial turning points by $r_{\rm p}$ and $r_{\tilde{a}}$, corresponding to the periapsis and apoapsis, respectively. At these points, the radial velocity vanishes, and the turning point condition reads $E^2=V_{\rm eff}$. Using the radial and azimuthal equations of motion, the total azimuthal angle accumulated during one radial period is

\begin{equation}
\label{eq:deltaphi}
\Delta\phi
=2\int_{r_{\rm p}}^{r_{\tilde{a}}}
\frac{L\,dr}
{(r^2+a^2)\sqrt{E^2-V_{\rm eff}}}.
\end{equation}

The factor of two accounts for the outward and inward portions of the radial motion. Following the classification of Ref.~\cite{levin2008periodic}, a periodic orbit is specified by three integers $(z,w,v)$, which describe its zoom, whirl, and vertex structure. Here, $z$ counts the radial oscillations required for the orbit to close, $w$ denotes the number of additional revolutions near periapsis, and $v$ determines the ordering of successive orbital lobes. These integers are related to the fundamental frequencies through

\begin{equation}
\label{eq:q1}
q=\frac{\omega_\phi}{\omega_r}-1
=w+\frac{v}{z},
\end{equation}

where $z\geq 1$, $w\geq 0$, and $0\leq v<z$. Since $\omega_\phi/\omega_r=\Delta\phi/(2\pi)$, the precession parameter can be expressed as

\begin{equation}
\label{eq:q2}
q=
\frac{1}{\pi}
\int_{r_{\rm p}}^{r_{\tilde{a}}}
\frac{L\,dr}
{(r^2+a^2)\sqrt{E^2-V_{\rm eff}}}-1.
\end{equation}

For a given spacetime, $q$ depends on the orbital energy and angular momentum, with regular core parameters via the effective potential term, which includes the metric functions.

To examine the influence of the regular core parameter $a$, we consider the angular momentum interval between the ISCO and the MBO, relevant to the zoom--whirl configurations studied here. We write

\begin{equation}
L=L_{\rm ISCO}
+\alpha\left(L_{\rm MBO}-L_{\rm ISCO}\right),
\end{equation}

with $0\leq\alpha\leq 1$. Throughout the following analysis, we choose $\alpha=0.5$, corresponding to

\begin{equation}
L_{av}=
\frac{L_{\rm ISCO}+L_{\rm MBO}}{2}.
\end{equation}

\begin{figure}[ht!]
    \centering
    \includegraphics[width=85mm]{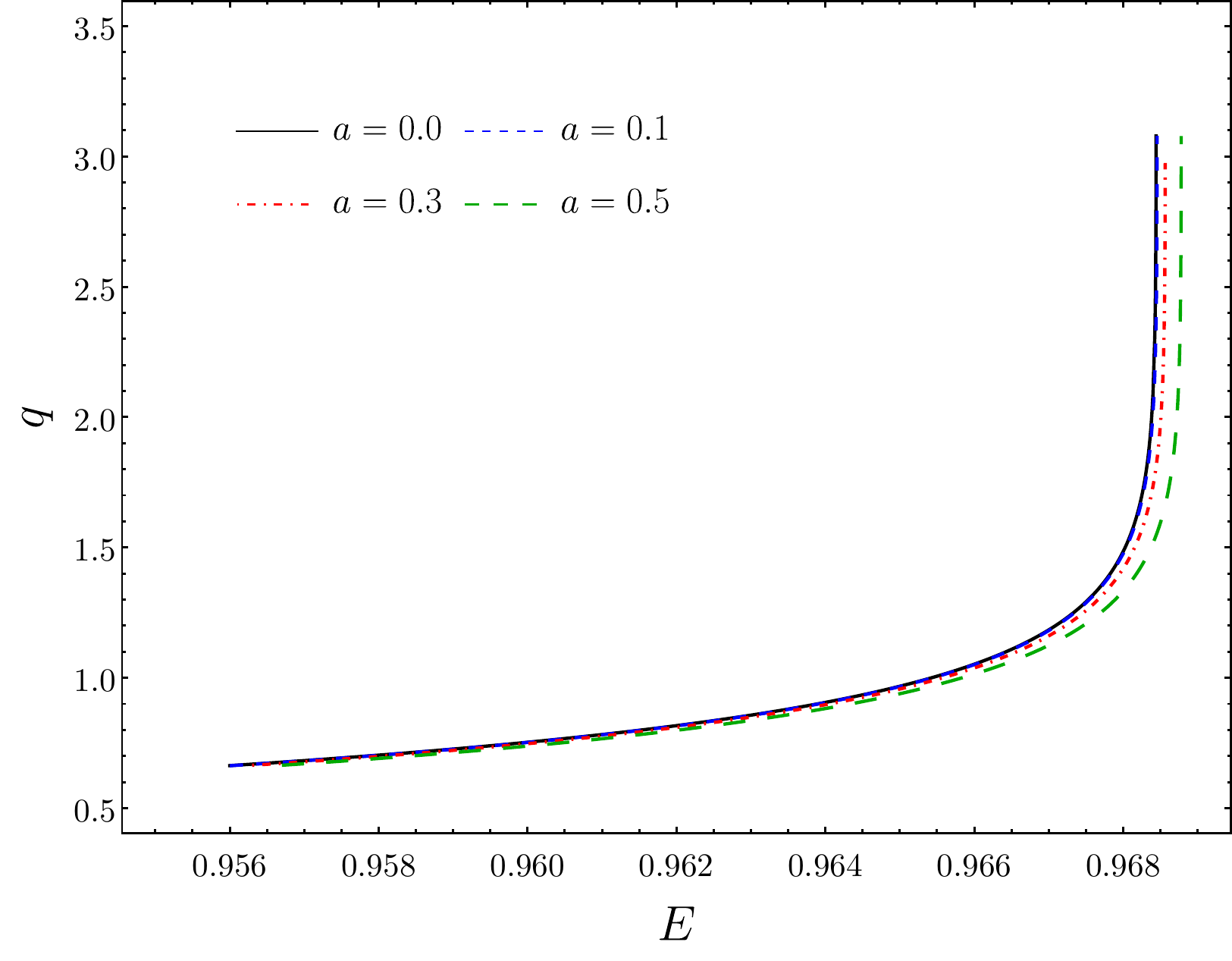}
    \caption{Precession parameter $q$ as a function of the orbital energy $E$ for different values of the regular core parameter $a$. For each $a$, the angular momentum is fixed at $L_{\rm av}$, with $M=1$.}
    \label{fig:qE}
\end{figure}

The angular momentum is therefore evaluated separately for each value of $a$. Fig.~\ref{fig:qE} shows the dependence of $q$ on the orbital energy for different values of the regular core parameter. In all cases, $q$ increases monotonically with $E$, with a much steeper rise toward the upper energy boundary, where the particle undergoes increasingly pronounced whirling motion. The curves nearly overlap at lower energies, indicating that the influence of the regular core is relatively weak in this regime. At higher energies, however, the differences become more apparent. Increasing $a$ shifts the curves toward larger energies, so that a slightly higher energy is required to obtain the same precession parameter. This behavior reflects the changes in the effective potential and orbital angular momentum associated with the DBI geometry.

The energies corresponding to representative periodic orbits are collected in Tab.~\ref{tab:qE1} for several values of $a$, with $M=1$. For each orbital configuration $(z,w,v)$, the required energy increases slightly as $a$ grows. For example, the energy of the $(3,1,2)$ orbit changes from $E=0.968225$ in the Schwarzschild limit to $E=0.968568$ at $a=0.5$. Over the same interval, $L_{av}$ increases from $3.73205$ to $3.74959$. Although these shifts are modest, they consistently indicate that the regular core parameter modifies the conditions under which a given periodic trajectory is realized.

\begin{figure}[htbp]
    \centering
    \begin{tabular}{@{}c@{\hspace{0.03\textwidth}}c@{\hspace{0.03\textwidth}}c@{}}
        \includegraphics[width=0.3\textwidth]{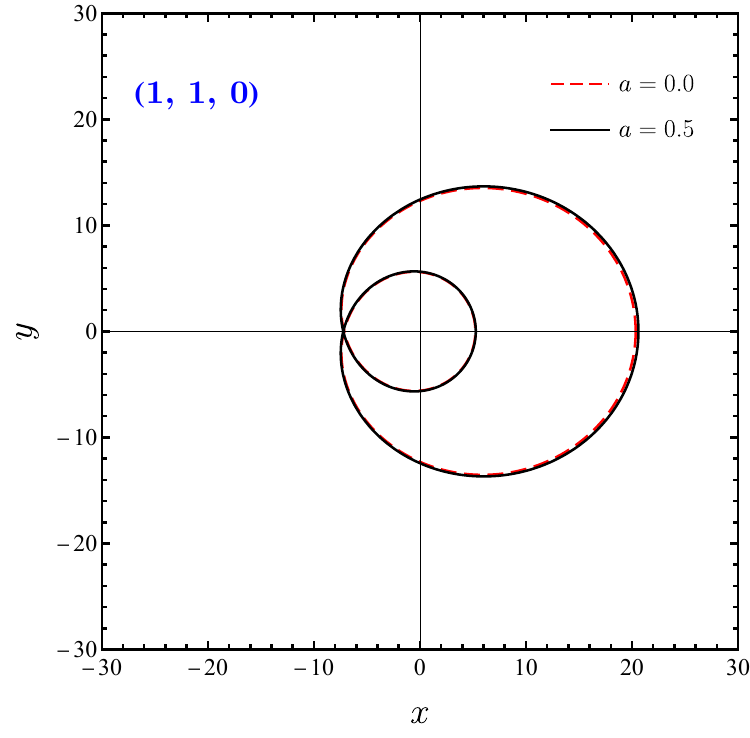} &
        \includegraphics[width=0.3\textwidth]{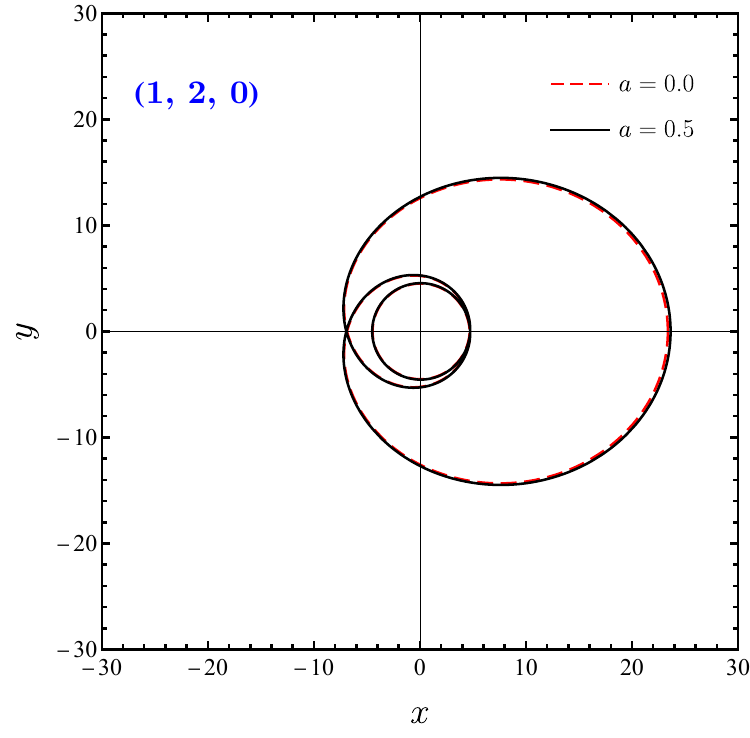} &
        \includegraphics[width=0.3\textwidth]{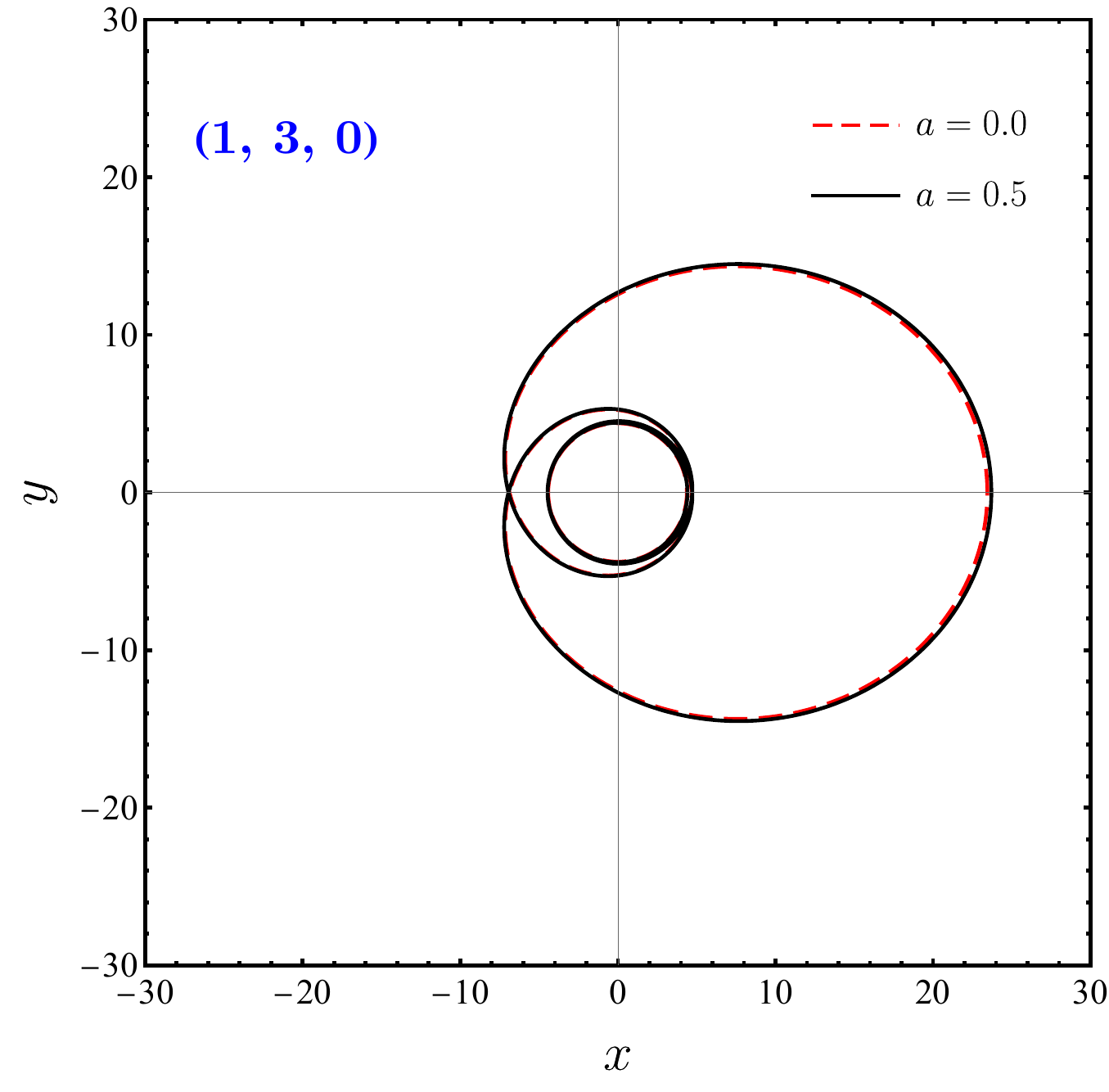} \\
        
        \includegraphics[width=0.3\textwidth]{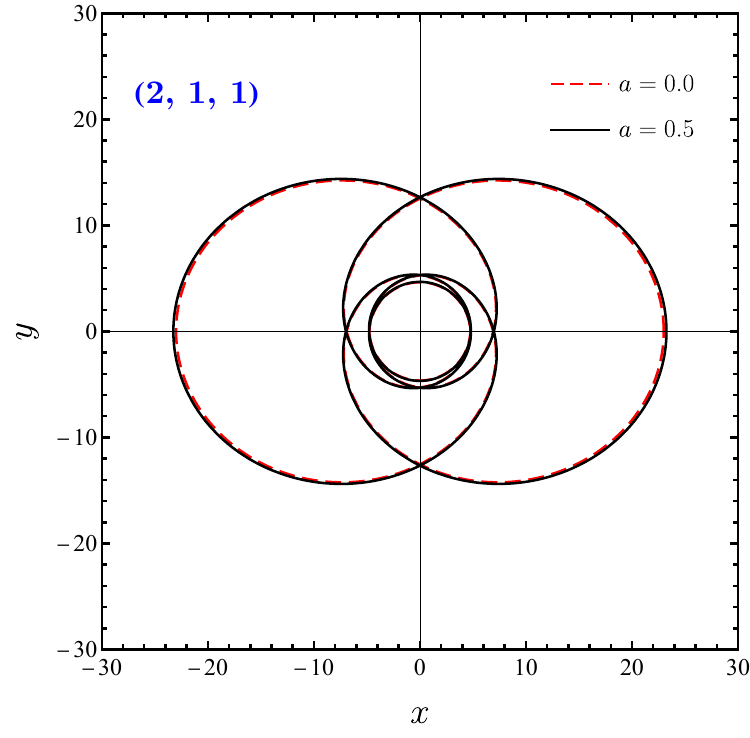} &
        \includegraphics[width=0.3\textwidth]{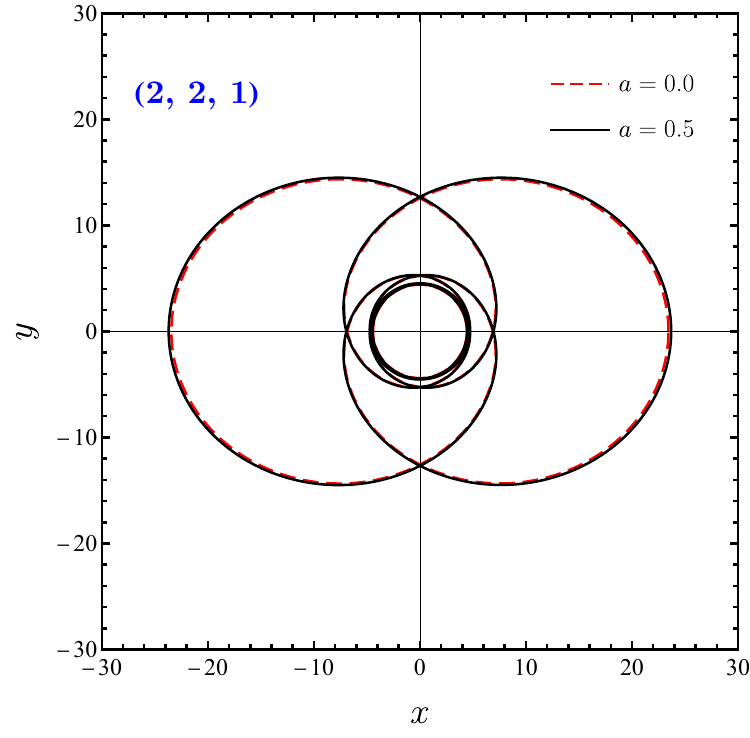} &
        \includegraphics[width=0.3\textwidth]{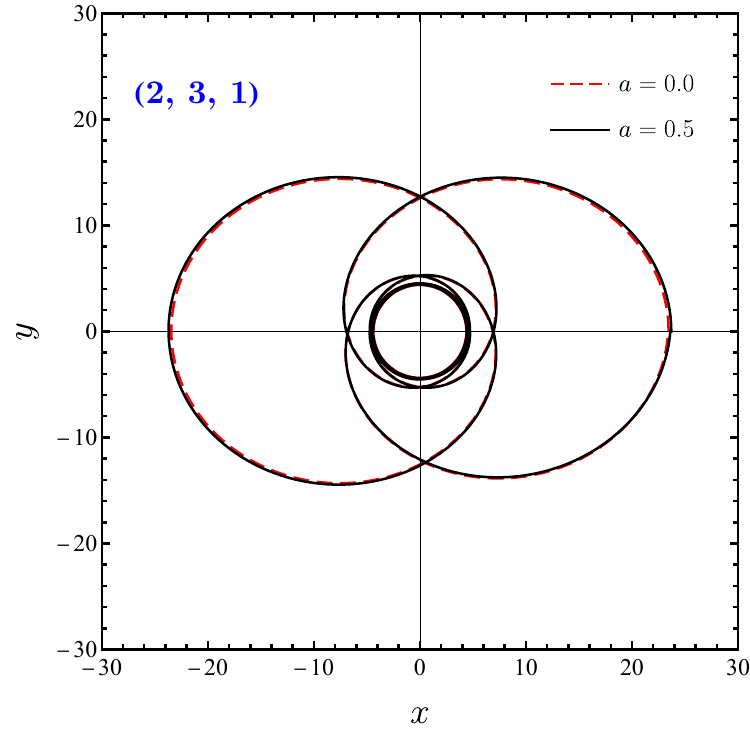} \\
        
         \includegraphics[width=0.3\textwidth]{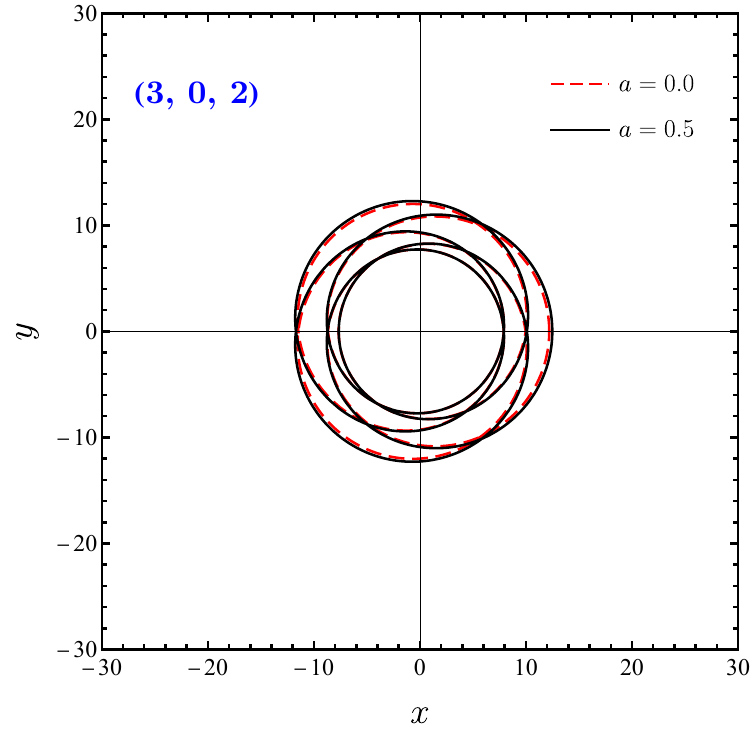} &
        \includegraphics[width=0.3\textwidth]{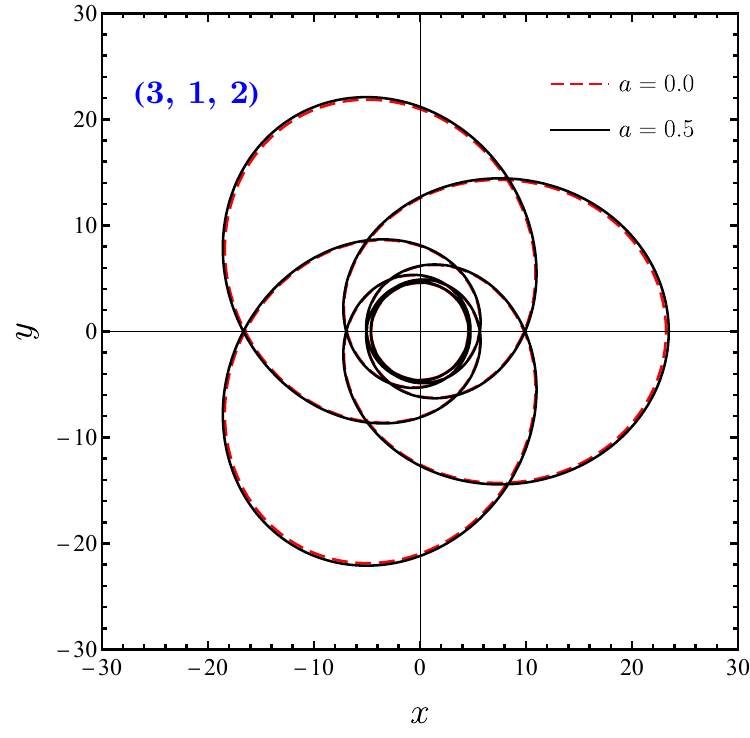} &
        \includegraphics[width=0.3\textwidth]{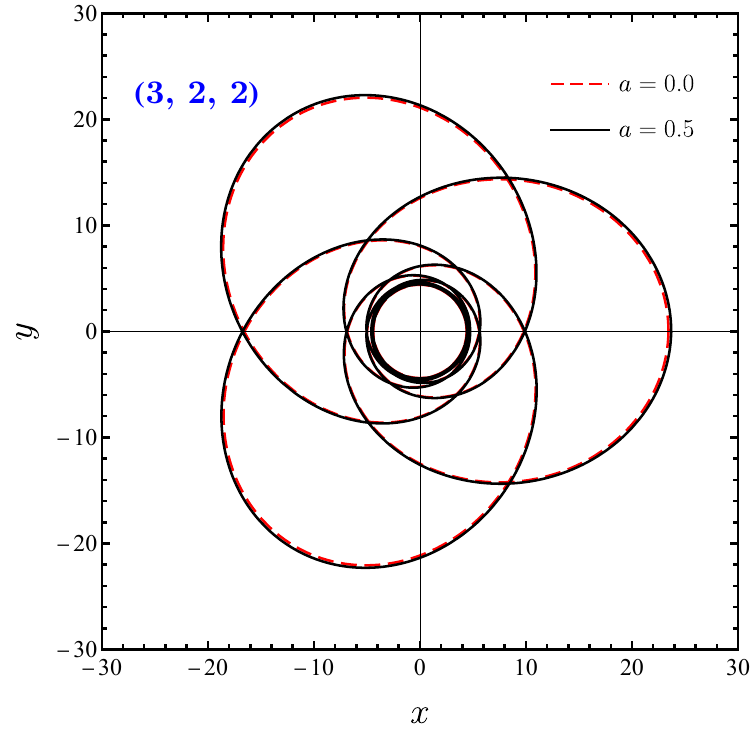} \\
        
         \includegraphics[width=0.3\textwidth]{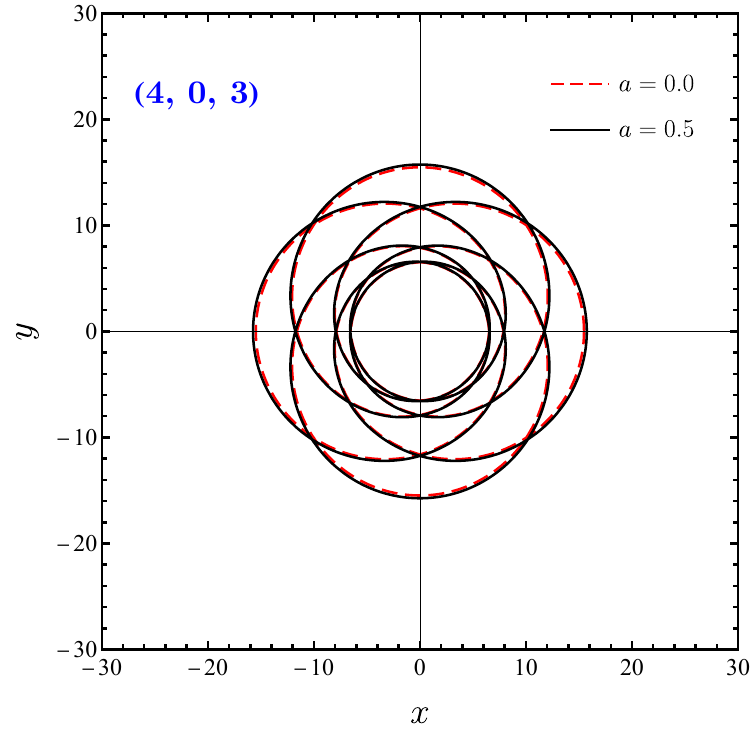} &
        \includegraphics[width=0.3\textwidth]{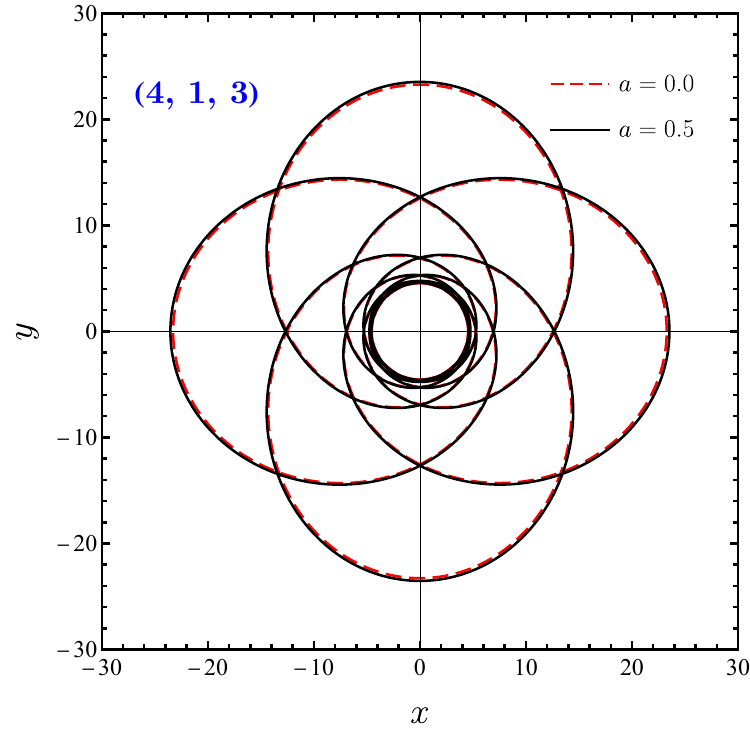} &
        \includegraphics[width=0.3\textwidth]{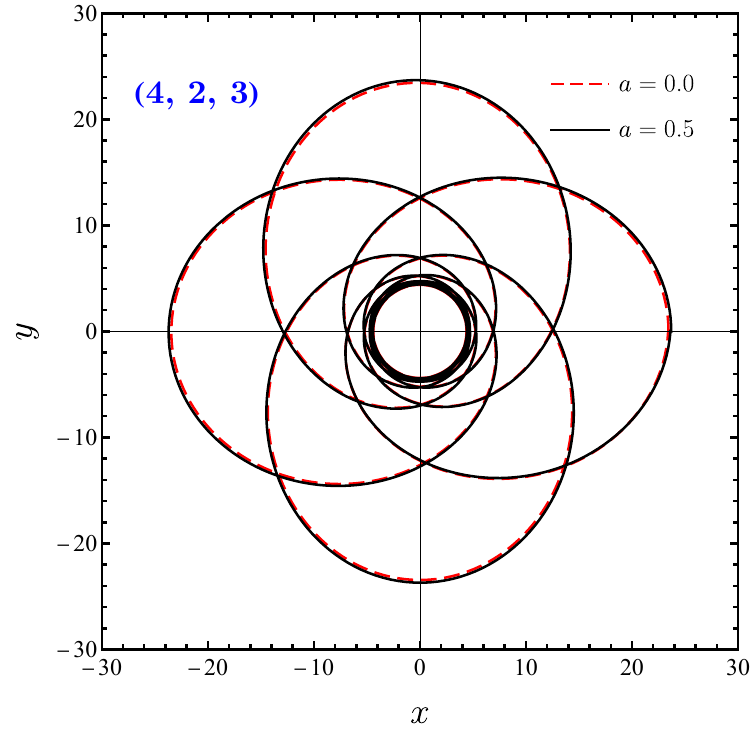} 
    \end{tabular}
    
  \caption{Comparison of representative periodic orbits around the Schwarzschild ($a=0$, red dashed curves) and phantom DBI regular ($a=0.5$, black solid curves) black holes. Each panel corresponds to the indicated $(z,w,v)$ configuration, with the energy and angular momentum chosen separately for each geometry. The mass is fixed at $M=1$.}
   
    \label{fig:allorbits}
\end{figure}

The corresponding orbital trajectories are shown in Fig.~\ref{fig:allorbits}, where the Schwarzschild and regular DBI geometries are compared for the same $(z,w,v)$ configurations. Increasing $z$ produces more intricate closed trajectories, while larger values of $w$ introduce additional revolutions near the central region. Although the two geometries exhibit identical zoom--whirl patterns, the DBI orbits are slightly more extended than their Schwarzschild counterparts. This small but noticeable difference indicates that the regular core parameter modifies the orbital geometry without significantly altering its overall structure.

\begin{table}[!ht]
\centering
\caption{Orbital energies $E$ for representative periodic orbits labeled by $(z,w,v)$ at different values of the regular core parameter $a$. The corresponding average angular momentum $L_{\rm av}$ is evaluated separately for each $a$, with $M=1$.}
\label{tab:qE1}
\small
\setlength{\tabcolsep}{4pt}
\begin{tabular}{|c|S[table-format=1.5]|*{8}{S[table-format=1.6]|}}
\hline
$a$ & {$L_{av}$} & {$E(1,1,0)$} & {$E(2,1,1)$} & {$E(3,1,2)$} & \
{$E(4,1,3)$} & {$E(1,2,0)$} & {$E(2,2,1)$} & {$E(3,2,2)$} & \
{$E(4,2,3)$} \\
\hline\hline
0.0 & 3.73205 & 0.965425 & 0.968026 & 0.968225 & 0.968285 & 0.968383 \
& 0.968434 & 0.968438 & 0.968440 \\
\hline
0.1 & 3.73276 & 0.965442 & 0.968041 & 0.968239 & 0.968299 & 0.968397 \
& 0.968448 & 0.968452 & 0.968454 \\
\hline
0.3 & 3.73842 & 0.965576 & 0.968154 & 0.968350 & 0.968410 & 0.968507 \
& 0.968558 & 0.968562 & 0.968563 \\
\hline
0.5 & 3.74959 & 0.965837 & 0.968375 & 0.968568 & 0.968627 & 0.968722 \
& 0.968772 & 0.968776 & 0.968777 \\
\hline
\end{tabular}
\end{table}

\section{Precession of periodic orbits}
\label{sec:precessing}

The closure of a periodic orbit requires a precise relation between the radial and azimuthal frequencies. A small departure from this condition generally prevents the trajectory from closing after the same number of radial cycles, causing successive apsides to gradually shift. To examine this behavior in the phantom DBI regular black hole spacetime, we consider small variations in the orbital energy while keeping the angular momentum fixed. This provides a direct way to assess how the orbital structure responds to deviations from the energies associated with exact periodic motion.

For a reference periodic orbit characterized by $(z,w,v)$, the precession parameter takes the rational value $q_0=w+v/z$. We introduce a small relative energy shift according to

\begin{equation}
E_{\rm pert}=E_0(1+\eta),
\qquad
\eta=\frac{\Delta E}{E_0},
\end{equation}

where $E_0$ denotes the energy of the exact periodic orbit. The corresponding change in the precession parameter is

\begin{equation}
\label{eq:deltaq}
\Delta q
=q(E_{\rm pert})-q(E_0).
\end{equation}

For a generic perturbation, the frequency ratio no longer satisfies the original closure condition, and the resulting orbit gradually shifts relative to its periodic counterpart. Since the azimuthal advance per radial cycle is $\Delta\phi=2\pi(1+q)$, the additional angular displacement introduced by the energy perturbation is $2\pi\Delta q$ per radial period.

\begin{figure}[t]
    \centering
    \includegraphics[width=0.48\textwidth]{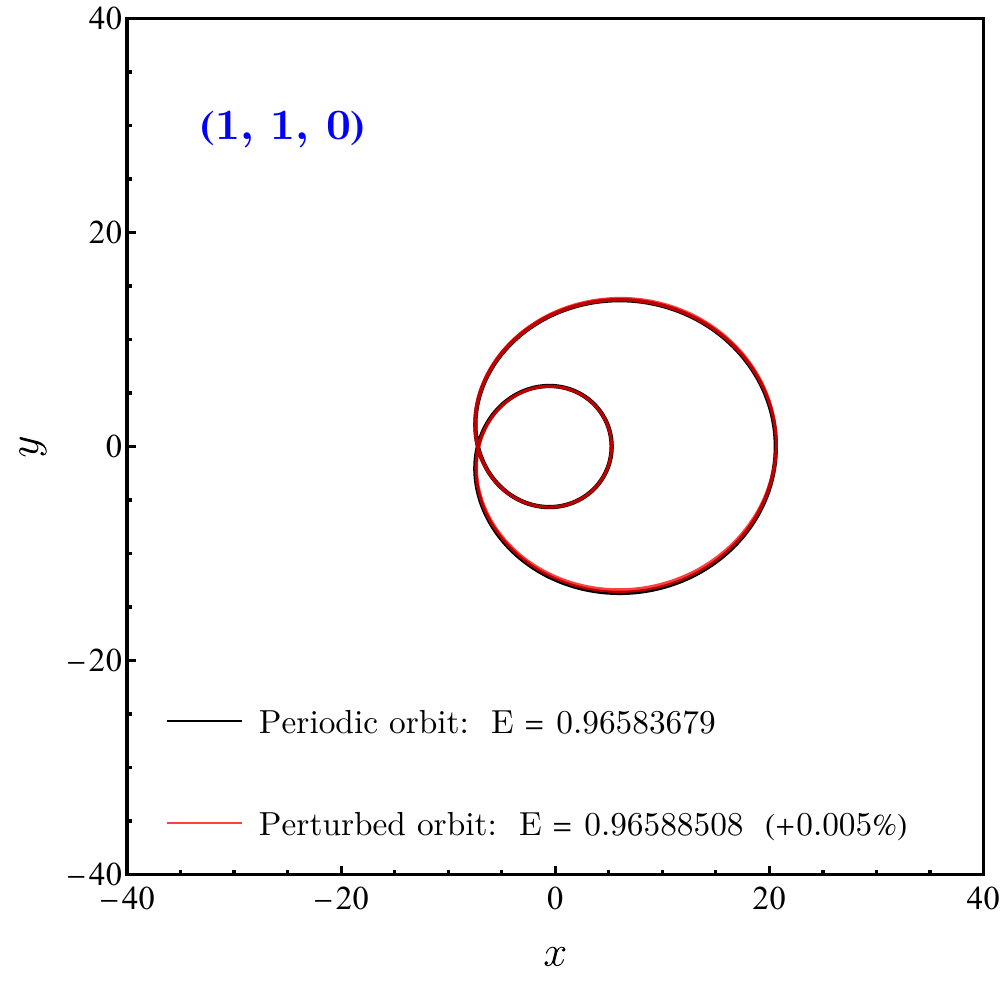}
    \hfill
    \includegraphics[width=0.48\textwidth]{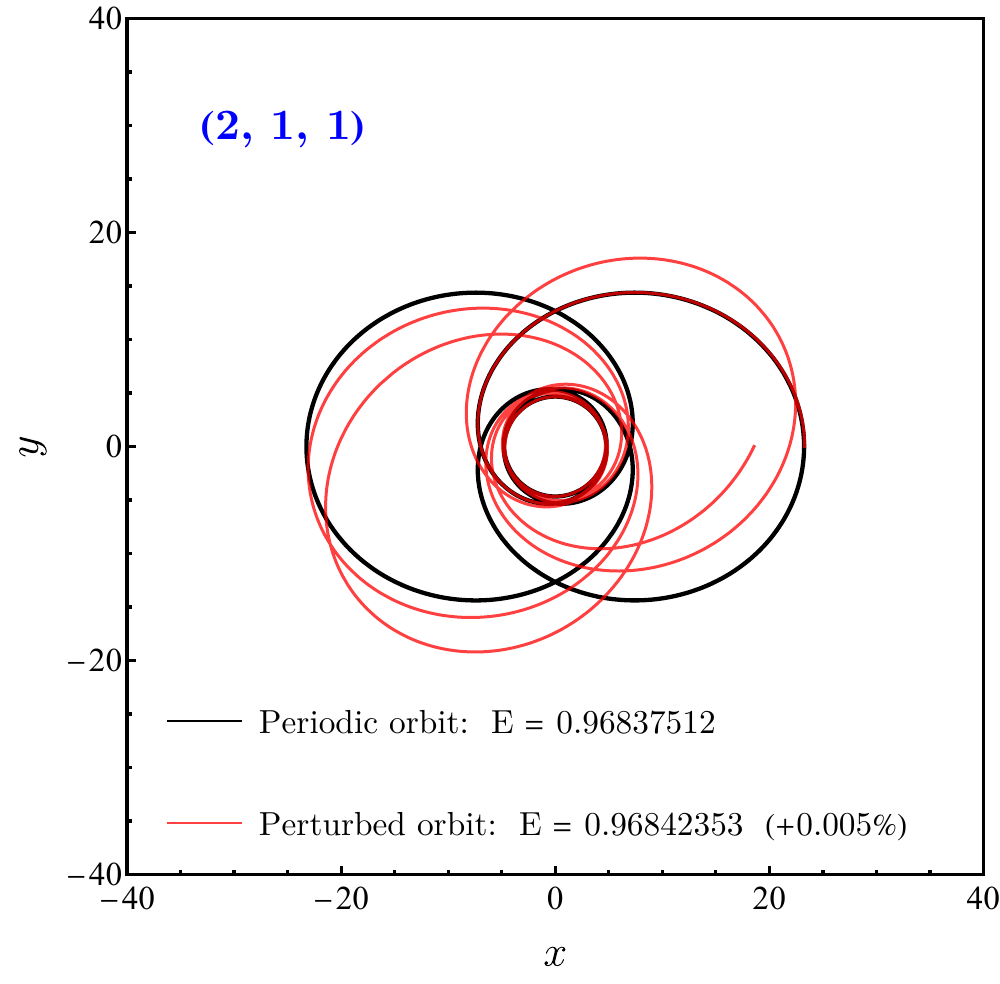}

    \medskip

    \includegraphics[width=0.48\textwidth]{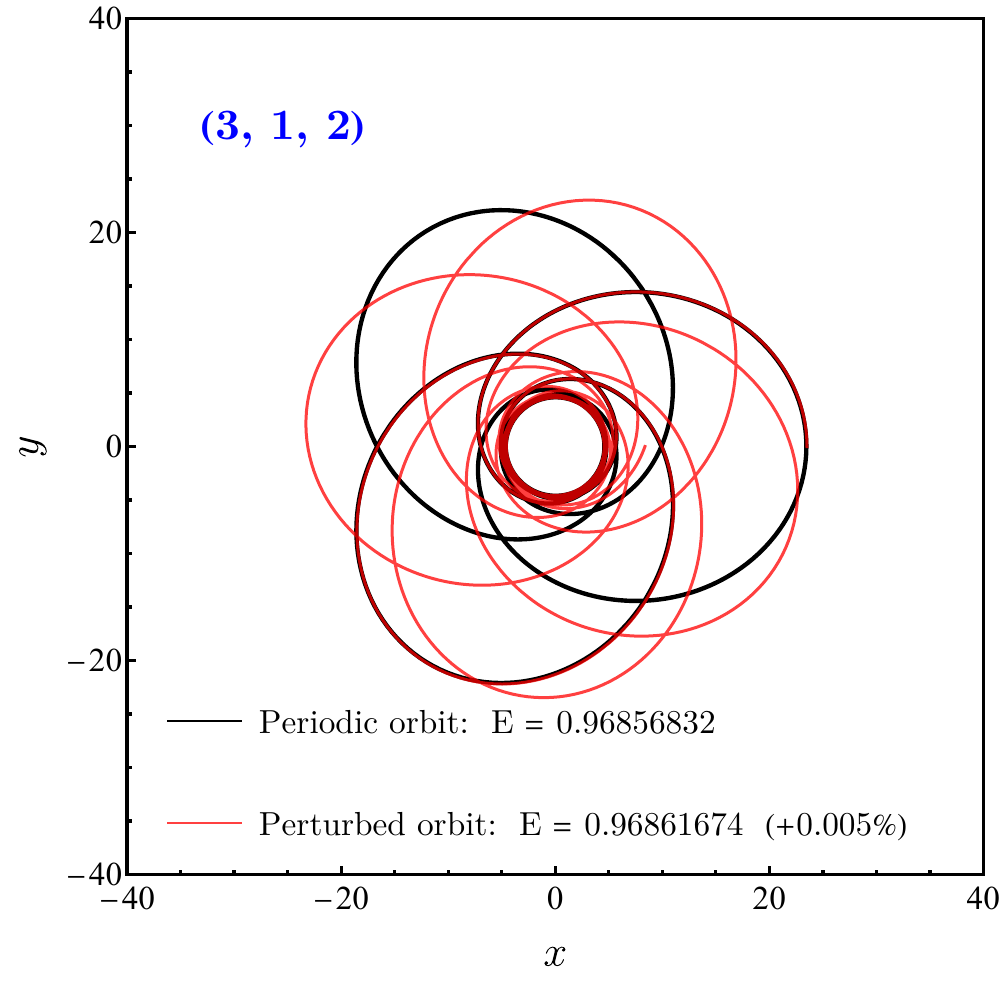}
    \hfill
    \includegraphics[width=0.48\textwidth]{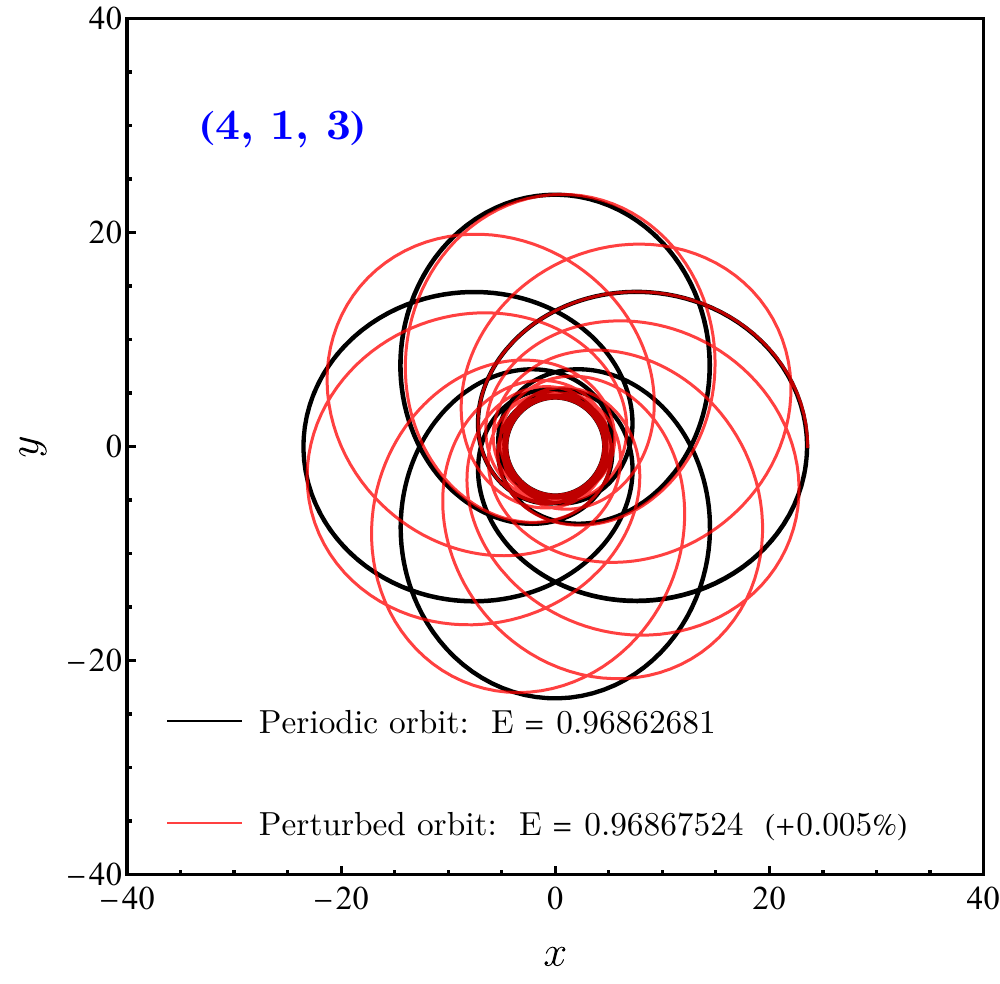}
    \caption{Comparison of exact periodic orbits (black) and their energy-perturbed counterparts (red) for four $(z,w,v)$ configurations. The calculations are performed with $a=0.5$, $M=1$, and $L=L_{\rm av}$. A relative energy increase of $\Delta E/E_0=0.005\%$ produces a departure from orbital closure.}
    \label{fig:pert}
\end{figure}

We illustrate this effect in Fig.~\ref{fig:pert} for four representative configurations, $(1,1,0)$, $(2,1,1)$, $(3,1,2)$, and $(4,1,3)$. The calculations are performed for $a=0.5$, with the angular momentum fixed at $L_{\rm av}$ and a common relative energy increase of $\eta=5\times10^{-5}$. The exact periodic trajectories are shown alongside their perturbed counterparts by black and red lines, respectively. Despite the small energy difference, the perturbed orbits no longer reproduce the original closed pattern over the displayed evolution. The departure is barely distinguishable for $(1,1,0)$, but becomes increasingly apparent for configurations with larger values of $q_0$, where the orbital lobes progressively shift in orientation.

The numerical results are summarized in Table~\ref{tab:delta}. Despite the identical relative energy perturbation applied to all four configurations, $\Delta q$ increases from approximately $0.0041$ for $(1,1,0)$ to $0.0980$ for $(4,1,3)$. This behavior reflects the increasing slope of $q(E)$ at higher reference energies, as shown in Fig.~\ref{fig:qE}. Consequently, the same relative energy change produces a more pronounced angular dephasing over successive radial cycles, revealing the sensitivity of near periodic motion to small energy variations in the phantom DBI spacetime.

\begin{table}[t]
\centering
\caption{Energies and precession parameters of selected periodic orbits and their perturbed counterparts for $a=0.5$ and $M=1$. The angular momentum is fixed at $L_{\rm av}$, and the same relative energy perturbation $\Delta E/E_0=5\times10^{-5}$ is applied to each configuration.}
\label{tab:delta}
\small
\setlength{\tabcolsep}{6pt}
\begin{tabular}{ccccccc}
\hline
$(z,w,v)$ & $E_0$ & $q_0$ &
$\Delta E/E_0$ & $E_{\rm pert}$ &
$q_{\rm pert}$ & $\Delta q$ \\
\hline\hline
$(1,1,0)$ & 0.965837 & 1.000000 &
0.00005 & 0.965885 & 1.004065 & 0.004065 \\
$(2,1,1)$ & 0.968375 & 1.500000 &
0.00005 & 0.968424 & 1.532668 & 0.032668 \\
$(3,1,2)$ & 0.968568 & 1.666667 &
0.00005 & 0.968617 & 1.733599 & 0.066932 \\
$(4,1,3)$ & 0.968627 & 1.750000 &
0.00005 & 0.968675 & 1.847992 & 0.097992 \\
\hline
\end{tabular}
\end{table}


\section{Constraints from stellar orbital precession}
\label{sec:precession}

The orbital motion of stars around Sgr~A$^*$ offers a means of testing the spacetime geometry in the vicinity of the Galactic centre. In particular, the GRAVITY Collaboration has measured the relativistic periastron advance of the S2 star, finding agreement with the Schwarzschild prediction within the observational uncertainties \cite{abuter2020detection,yao2026probing, li2024precessing,lin2022precessing}. This measurement provides an opportunity to examine whether the regular core parameter of the DBI black hole can be constrained by stellar dynamics.

We consider timelike geodesics in the equatorial plane and retain the radial coordinate $r$ throughout the calculation, consistently accounting for the modified angular sector of the metric.

For a bound orbit, the radial motion can be parametrized in terms of the relativistic anomaly $\psi$ as
\begin{equation}
r(\psi)=\frac{p}{1+e\cos\psi},
\qquad
p=\tilde{a}(1-e^2),
\label{eq:prec_radial}
\end{equation}
where $e$ is the orbital eccentricity and $\tilde{a}$ denotes the semi-major axis. The radial turning points are then
\begin{equation}
r_p=\tilde{a}(1-e),
\qquad
r_a=\tilde{a}(1+e),
\label{eq:prec_turning}
\end{equation}
corresponding to periastron and apastron, respectively. Here, $\tilde{a}$ and $e$ are defined with respect to the coordinate $r$; their relation to the corresponding areal-radius quantities approaches the usual Keplerian form in the weak--field limit. 
The orbital precession can be evaluated from the apsidal angle derived in Eq.~\eqref{eq:deltaphi}. To express this quantity in terms of the relativistic anomaly $\psi$, we use the radial parametrization introduced in Eq.~\eqref{eq:prec_radial}. The conserved quantities $E$ and $L$ are determined by imposing the turning-point conditions
\begin{equation}
E^2=V_{\rm eff}(r_p)=V_{\rm eff}(r_{\tilde{a}}).
\label{eq:prec_turning_condition}
\end{equation}
For the DBI geometry, these conditions yield
\begin{equation}
E^2=
\frac{f(r_p)f(r_{\tilde{a}})(r_{\tilde{a}}^2-r_p^2)}{f(r_p)(r_{\tilde{a}}^2+a^2)
-f(r_{\tilde{a}})(r_p^2+a^2)},
\label{eq:prec_E}
\end{equation}
and
\begin{equation}
L^2=
\frac{
(r_p^2+a^2)(r_{\tilde{a}}^2+a^2)
[f(r_{\tilde{a}})-f(r_p)]}{f(r_p)(r_{\tilde{a}}^2+a^2)
-f(r_{\tilde{a}})(r_p^2+a^2)}.
\label{eq:prec_L}
\end{equation}

Combining the radial parametrization with the geodesic equations, the azimuthal evolution along the outward branch of the orbit becomes
\begin{equation}
\frac{d\phi}{d\psi}
=
\frac{Le\sin\psi}
{p\sqrt{
E^2-f(r(\psi))
\left[
1+\dfrac{L^2}{r^2(\psi)+a^2}
\right]
}}
\frac{r^2(\psi)}{r^2(\psi)+a^2},
\label{eq:prec_dphi_dpsi}
\end{equation}
where $0<\psi<\pi$. The factor
$r^2/(r^2+a^2)$ arises from the angular part of the DBI metric and distinguishes the present expression from its Schwarzschild counterpart.

The total azimuthal angle accumulated during one radial period can therefore be expressed as
\begin{equation}
\Delta\phi
=
2\int_0^\pi
\frac{d\phi}{d\psi}\,d\psi,
\label{eq:prec_total_angle}
\end{equation}
with the corresponding periastron advance defined by
\begin{equation}
\Delta\omega_{\rm DBI}
=
\Delta\phi-2\pi.
\label{eq:prec_advance}
\end{equation}

To obtain an analytical expression for the periastron advance, we consider the weak--field regime, where the orbital scale is much larger than both the gravitational radius and the regular-core parameter,
\begin{equation}
\frac{M}{p}\ll 1,
\qquad
\frac{a^2}{p^2}\ll 1.
\label{eq:prec_weakfield}
\end{equation}
In this limit, the metric function admits the asymptotic expansion. Substituting the turning-point expressions for $E$ and $L$ into Eq.~\eqref{eq:prec_dphi_dpsi}, we expand the azimuthal evolution to first order in $M/p$ and $a^2/p^2$. Integrating over a complete radial period gives
\begin{equation}
\Delta\phi \simeq
2\pi+\frac{6\pi M}{p}
+\frac{\pi a^2}{p^2}
\left(\frac{14}{5}+\frac{e^2}{2}\right),
\label{eq:prec_delta_phi_exp}
\end{equation}
where terms of higher order in the weak--field parameters have been neglected.

Subtracting the Newtonian contribution, the periastron advance becomes
\begin{equation}
\Delta\omega_{\rm DBI}
\simeq
\frac{6\pi M}{p}
+\frac{\pi a^2}{p^2}
\left(\frac{14}{5}+\frac{e^2}{2}\right).
\label{eq:prec_DBI}
\end{equation}
Using $p=\tilde{a}(1-e^2)$, this can be expressed in terms of the orbital elements as
\begin{equation}
\Delta\omega_{\rm DBI}
\simeq
\frac{6\pi M}{\tilde{a}(1-e^2)}
+
\frac{\pi a^2}{\tilde{a}^2(1-e^2)^2}
\left(\frac{14}{5}+\frac{e^2}{2}\right).
\label{eq:prec_DBI_orbital}
\end{equation}

The first term reproduces the familiar Schwarzschild precession, while the second accounts for the leading modification introduced by the DBI regular core. The correction is positive and grows quadratically with $a$, indicating an enhanced prograde precession relative to the Schwarzschild case. Its dependence on the orbital scale also shows that the deviation becomes progressively smaller for more distant orbits.

To quantify this deviation, we introduce the ratio of the predicted periastron advance to its Schwarzschild value,
\begin{equation}
f_{sp}^{\rm DBI}
\equiv
\frac{\Delta\omega_{\rm DBI}}
{\Delta\omega_{\rm GR}},
\label{eq:prec_ratio_def}
\end{equation}
where
\begin{equation}
\Delta\omega_{\rm GR}
=\frac{6\pi M}{\tilde{a}(1-e^2)}.
\end{equation}
At leading order, we obtain
\begin{equation}
f_{sp}^{\rm DBI}
\simeq
1+
\left(\frac{7}{15}+\frac{e^2}{12}\right)
\frac{a^2}{M \tilde{a}(1-e^2)}.
\label{eq:precratio}
\end{equation}

\begin{figure}[ht!]
\centering
\includegraphics[width=85mm]{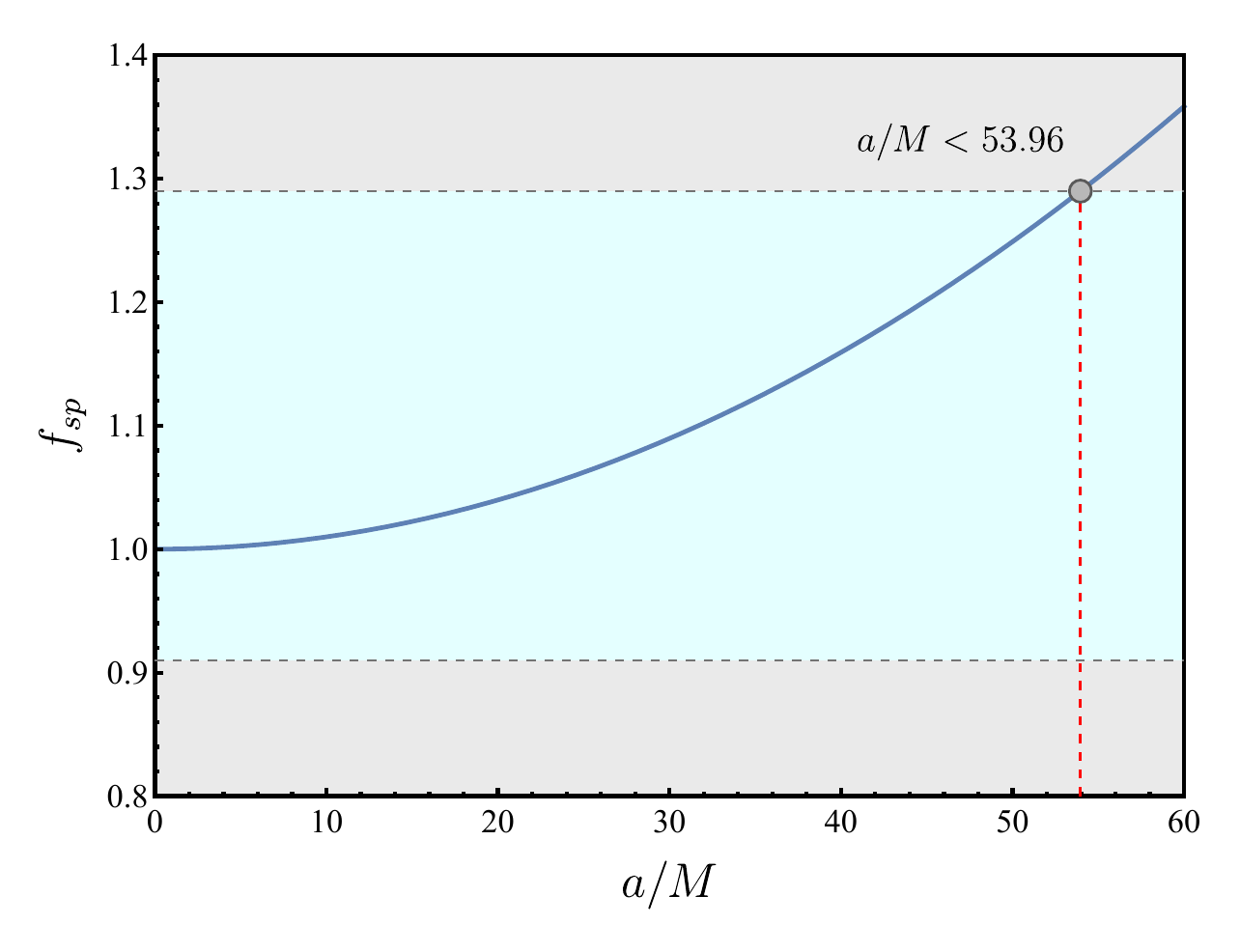}
\caption{Dependence of the precession ratio $f_{sp}$ on the dimensionless parameter $a/M$. The blue shaded band represents the allowed region inferred from the measured periastron precession of the S2 star.}
\label{fig:S2}
\end{figure}

The Schwarzschild limit is recovered at $a=0$. For nonzero $a$, the precession ratio exceeds unity, although the correction is suppressed by the small factor $1/[M\tilde{a}(1-e^2)]$. The size of this effect can therefore be assessed by comparing Eq.~\eqref{eq:precratio} with the measured precession of the S2 star. We use the best-fit orbital parameters reported by the GRAVITY Collaboration \cite{abuter2020detection}. The parameters are an eccentricity $e=0.884649$, an angular $semi$-$major$ axis of $125.058\,\mathrm{mas}$, a distance from the Galactic center of $R_0=8246.7\,\mathrm{pc}$ and a central black hole mass of $M=4.261\times10^6 M_\odot$. The corresponding dimensionless semi-latus rectum is $p/M\approx5.34\times10^3$. In Fig.~\ref{fig:S2}, we calculate the predicted precession ratio $f_{sp}^{\rm DBI}$ using these values and compare it to the GRAVITY measurement $f_{sp}=1.10\pm0.19$, as a function of the dimensionless regular core parameter $a/M$, which results in the following constraint

\begin{equation}
a<53.9587M.
\end{equation}
The S2 precession measurement yields a formal upper limit of $a/M<53.96$ within the adopted weak--field approximation. However, the existence of a nonextremal event horizon already requires $a/M<3\pi/2\simeq4.71$. The observational bound is therefore substantially weaker than the theoretical restriction and does not further constrain the black hole configurations considered here.

\section{Gravitational waveforms from periodic orbits}
\label{sec:GW}

We now examine the gravitational wave signals associated with periodic motion around the phantom DBI regular black hole. We consider a stellar-mass compact object orbiting a supermassive black hole in the extreme mass-ratio regime, providing a representative configuration for studying gravitational radiation from EMRI systems. Such signals carry information about the underlying orbital dynamics and may reveal signatures of deviations from the Schwarzschild geometry \cite{cardoso2022gravitational,rahman2024probing}. We work within the adiabatic approximation, in which the orbital evolution driven by gravitational wave emission occurs on a timescale much longer than the orbital period \cite{hughes2000evolution,hughes2001evolution}. Accordingly, the particle's energy and angular momentum are treated as constant over the interval considered, and radiation-reaction effects are neglected when computing the waveforms along the periodic geodesics.


To construct the time-domain waveforms, we employ the numerical-kludge method~\cite{babak2007kludge}, which combines relativistic geodesic motion with the leading-order quadrupole approximation for gravitational radiation. The particle trajectory is first determined by numerically integrating the equations of motion, Eqs.~\eqref{eq:t_equation}--\eqref{eq:radial_equation}. The resulting motion is then mapped onto an auxiliary flat-space trajectory, allowing the gravitational wave polarizations to be evaluated from the mass quadrupole moment.

For a compact object of mass $m$ orbiting a black hole of mass $M$, the symmetric trace-free (STF) mass quadrupole moment is given by

\begin{equation}
\label{eq:quadrupole-moment}
I_{ij}=\mu
\left(X_iX_j-\frac{1}{3}\delta_{ij}X_kX^k\right),
\end{equation}
where $\mu=mM/(m+M)$ is the reduced mass. Following the numerical kludge prescription, we map the equatorial trajectory onto an auxiliary Cartesian coordinate system through $X=r\cos\phi$, $Y=r\sin\phi$, and $Z=0$. Using the Newtonian central-acceleration term in the quadrupole calculation, the far field metric perturbation takes the approximate form~\cite{zhao2025periodic}

\begin{equation}
\label{eq:quadrupole}
h_{ij}^{\rm TT}
\simeq \frac{4\nu M}{D_L}
\left(v_i v_j-\frac{M}{r}n_i n_j\right)^{\rm TT},
\end{equation}

where $\nu=mM/(M+m)^2$ is the symmetric mass ratio, $D_L$ is the luminosity distance, and $n_i$ and $v_i$ denote the components of the unit position vector and orbital velocity, respectively. To extract the two gravitational wave polarizations, we introduce an observer adapted orthonormal basis~\cite{poisson2014gravity,zhao2025periodic,yang2025gravitational},

\begin{align}
\mathbf{e}_X &=
(\cos\zeta,-\sin\zeta,0), \nonumber\\
\mathbf{e}_Y &=
(\cos\iota\sin\zeta,\cos\iota\cos\zeta,-\sin\iota),
\nonumber\\
\mathbf{e}_Z &=
(\sin\iota\sin\zeta,\sin\iota\cos\zeta,\cos\iota),
\label{eq:basis}
\end{align}

where $\iota$ denotes the inclination between the orbital angular momentum and the observer's line of sight, while $\zeta$ specifies the orientation within the orbital plane. The plus and cross polarizations are obtained by projecting the metric perturbation onto this basis,

\begin{equation}
h_+=\frac{1}{2}
(e_X^ie_X^j-e_Y^ie_Y^j)h_{ij},
\qquad
h_\times=e_X^ie_Y^jh_{ij}.
\end{equation}

Defining $\Psi=\phi+\zeta$, together with the radial and tangential velocities $\dot r=dr/dt$ and $v_\phi=r\,d\phi/dt$, we obtain

\begin{align}
h_+ = &\frac{\nu M}{D_L}
\Bigg[
\sin^2\iota
\left(\dot r^2+v_\phi^2-\frac{M}{r}\right)+(1+\cos^2\iota)
\left(\dot r^2-v_\phi^2-\frac{M}{r}\right)
\cos(2\Psi)\nonumber\\
&
-2(1+\cos^2\iota)
\dot r v_\phi\sin(2\Psi)
\Bigg],
\label{eq:hplus}
\end{align}

and

\begin{align}
h_\times =
\frac{2\nu M}{D_L}\cos\iota
\Bigg[
&\left(\dot r^2-v_\phi^2-\frac{M}{r}\right)
\sin(2\Psi)+2\dot r v_\phi\cos(2\Psi)
\Bigg].
\label{eq:hcross}
\end{align}

These expressions incorporate both radial and azimuthal velocity contributions, allowing the characteristic features of noncircular zoom--whirl motion to be reflected in the waveforms.

The geodesic equations are solved numerically in units of the black hole mass, with $M=1$. For the gravitational wave analysis, we consider a black hole of physical mass $10^7M_\odot$ and a symmetric mass ratio $\nu=10^{-6}$. The source is located at a luminosity distance of $D_L=200\,\mathrm{Mpc}$, corresponding to $D_L/M\simeq4.18\times10^{14}$. We fix the inclination and orientation angles at $\iota=\pi/4$ and $\zeta=0$, respectively.

\begin{figure}[ht!]
    \centering
    \begin{tabular}[t]{cc}
        \includegraphics[height=65mm]{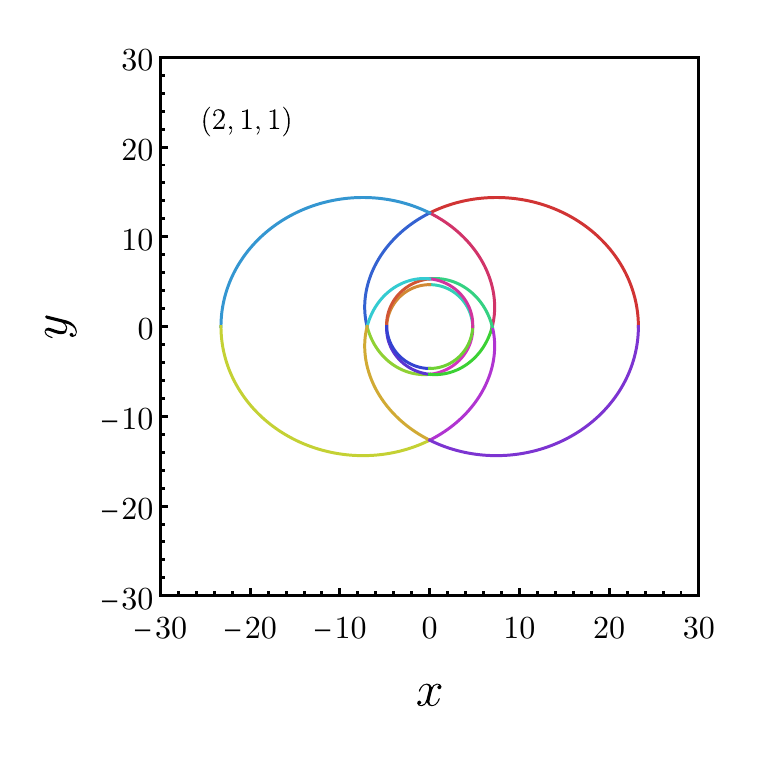}
       \includegraphics[height=65mm]{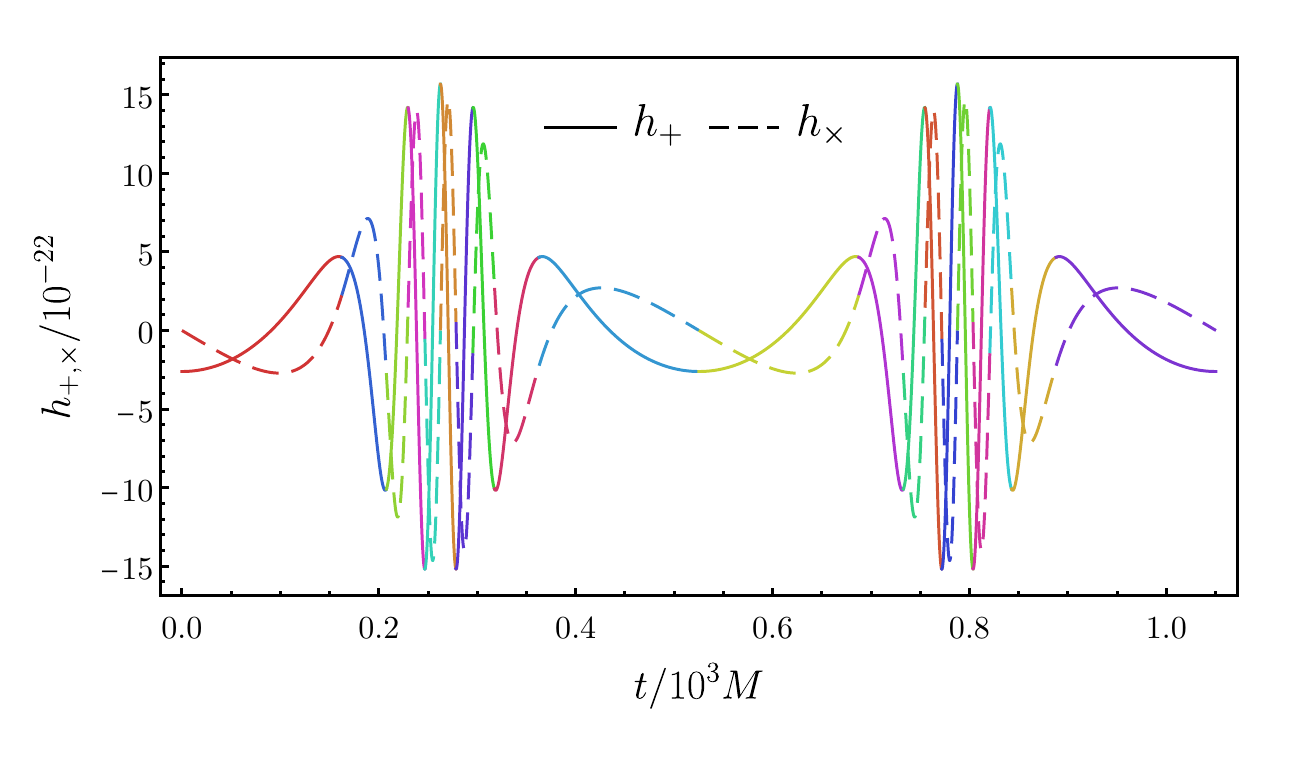} 
    \end{tabular}
    \caption{The $(2,1,1)$ periodic orbit (left) and its associated gravitational wave polarizations (right) for $a=0.5$ and $M=1$. Matching colors identify successive
$\pi/2$ intervals of the orbital phase, with solid and dashed curves representing $h_+$ and $h_\times$, respectively.}
   \label{figs:colored}
\end{figure}

\begin{figure}[ht!]
    \centering
    \includegraphics[width=55mm]{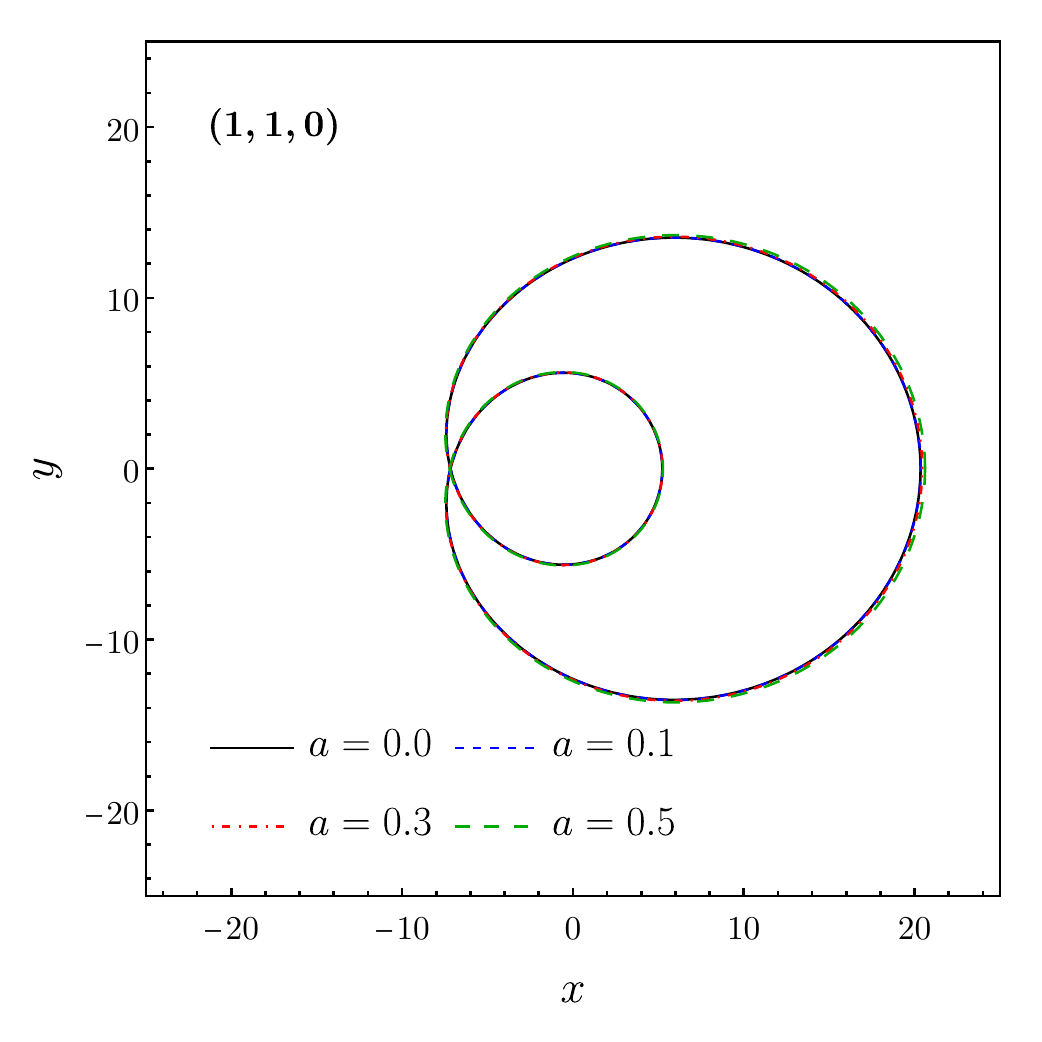}
    \includegraphics[width=55mm]{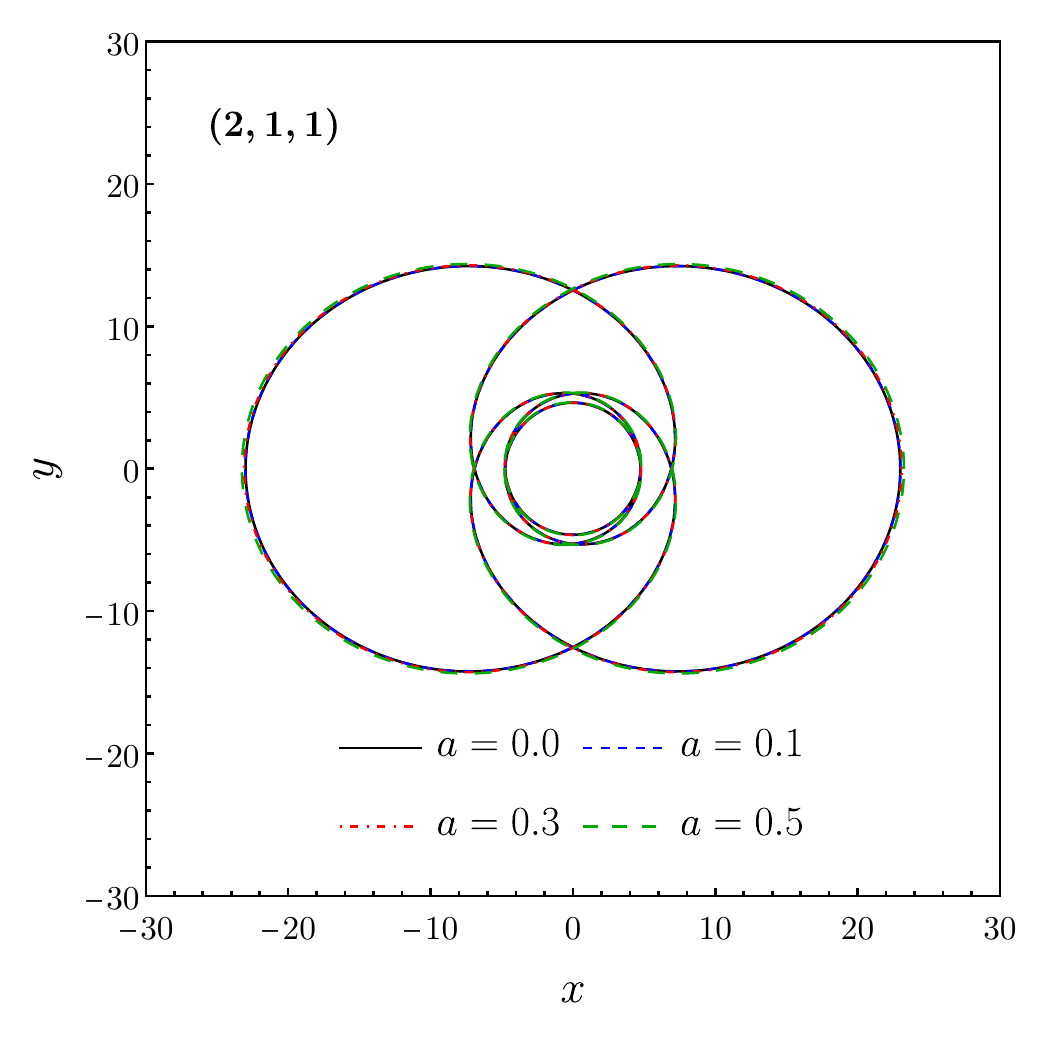}
    \includegraphics[width=55mm]{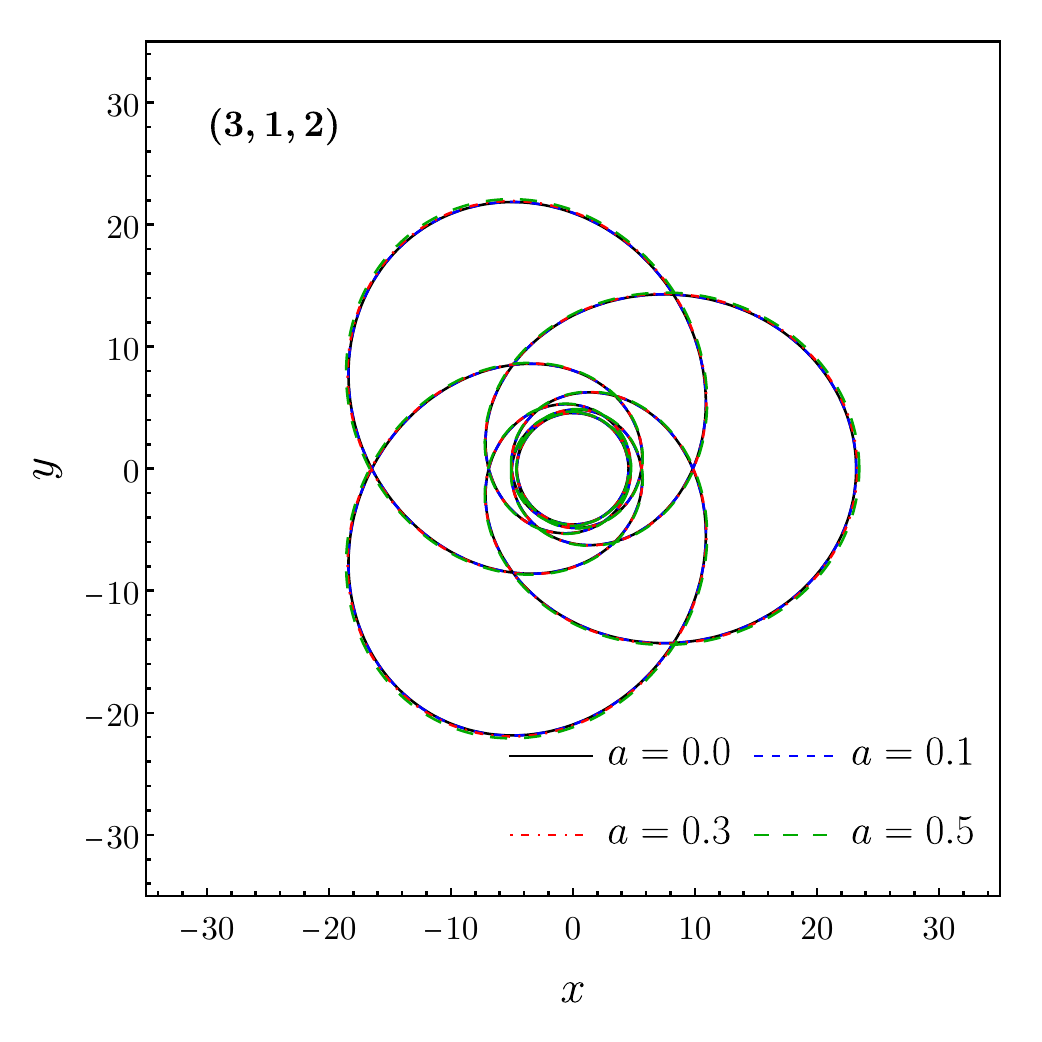}
    \caption{Periodic trajectories $(1,1,0)$, $(2,1,1)$,
and $(3,1,2)$ for different values of the regular
core parameter $a$, with $M=1$.}.
    \label{fig:comporbit}
\end{figure}

\begin{figure}[ht!]
    \centering
   \includegraphics[width=150mm]{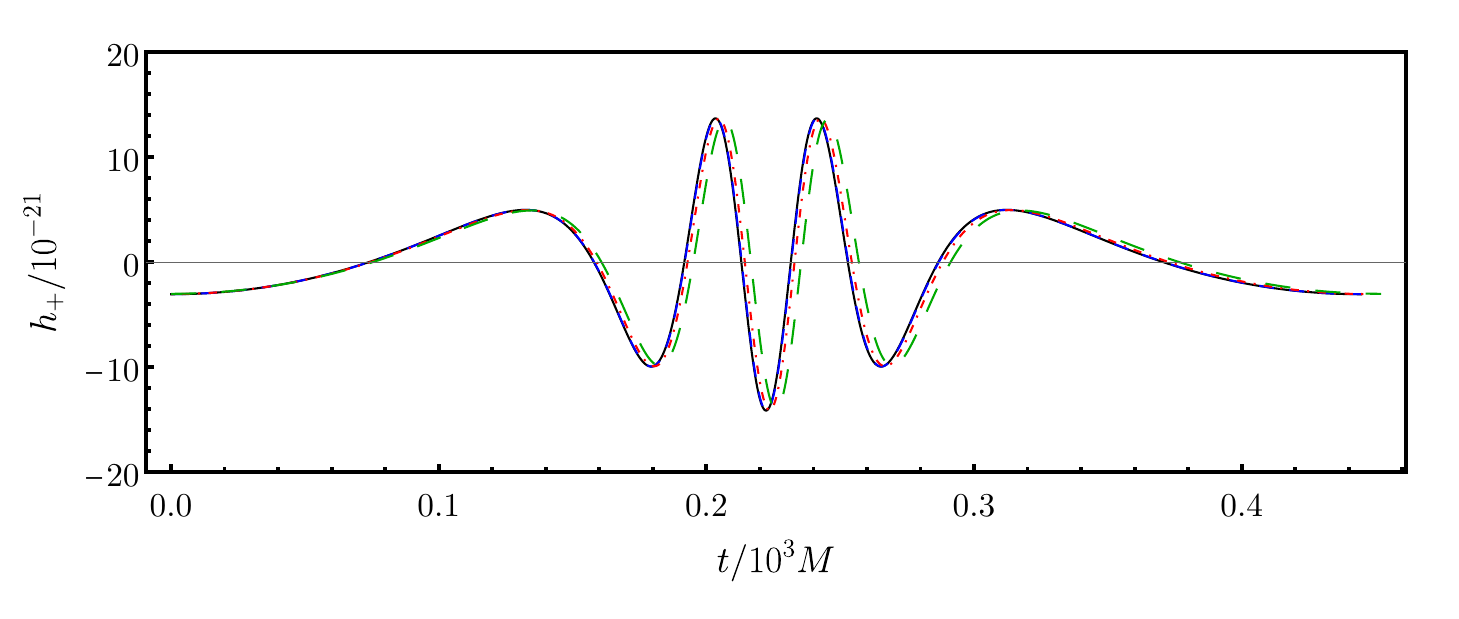}
    \includegraphics[width=150mm]{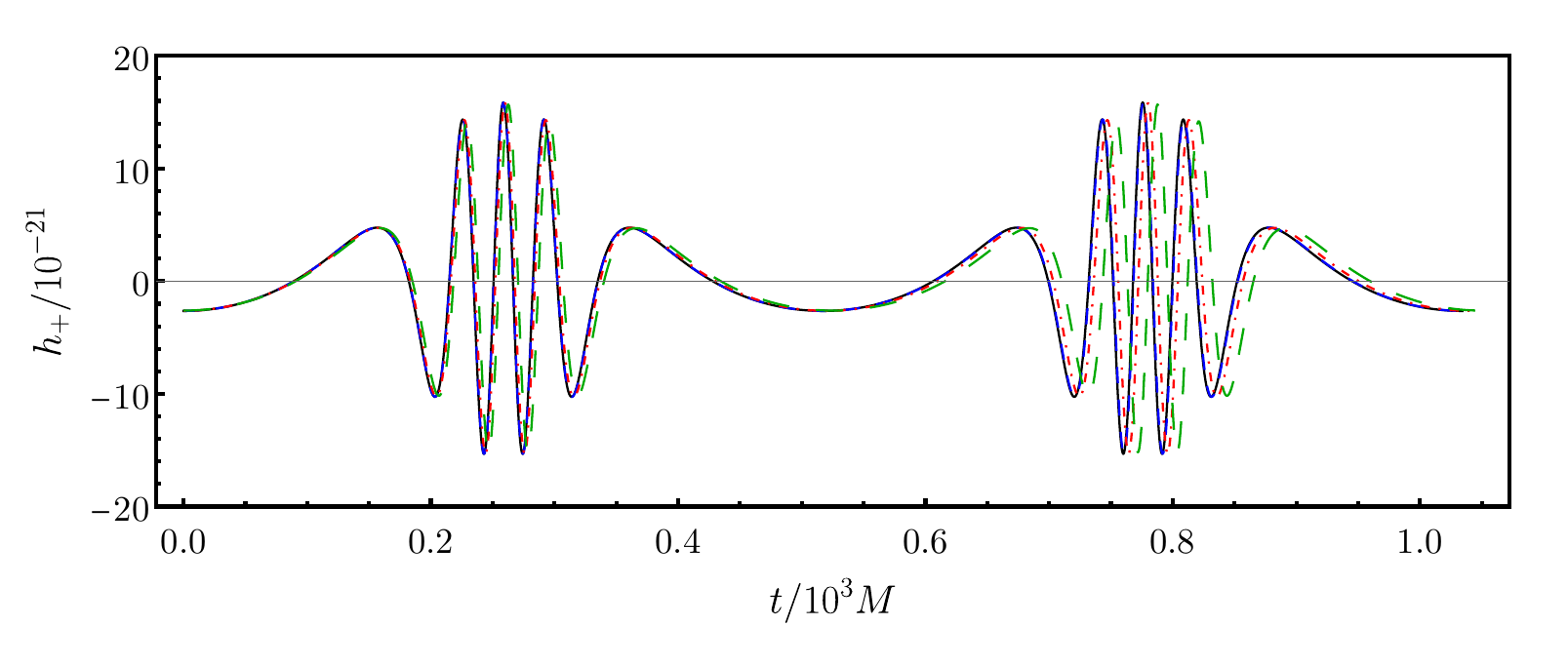}
    \includegraphics[width=150mm]{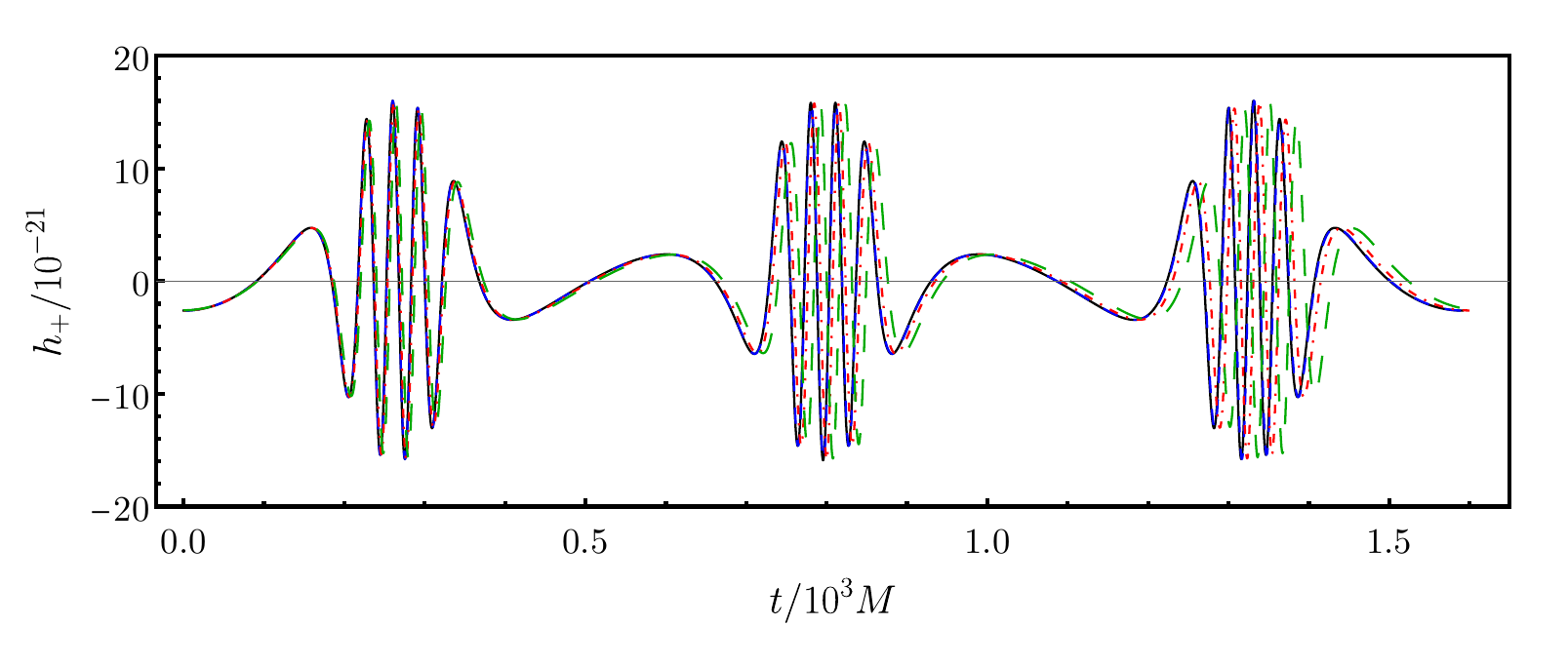}
     \caption{Plus polarization $h_+$ for the periodic orbits $(1,1,0)$, $(2,1,1)$, and $(3,1,2)$. The curves correspond to $a=0$ (black solid), $0.1$ (blue dashed), $0.3$ (red dot-dashed), and $0.5$ (green dashed).}
    \label{fig:comphplus}
\end{figure}

\begin{figure}[ht!]
    \centering
    \includegraphics[width=150mm]{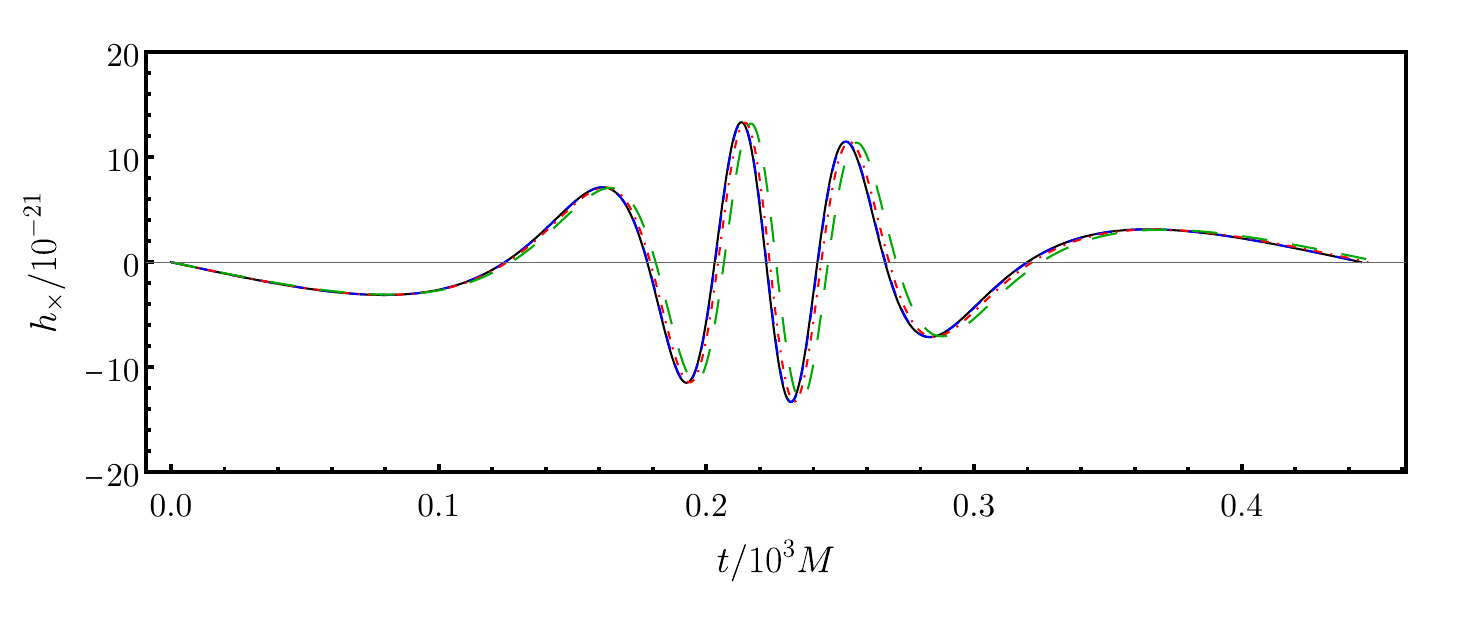}
   \includegraphics[width=150mm]{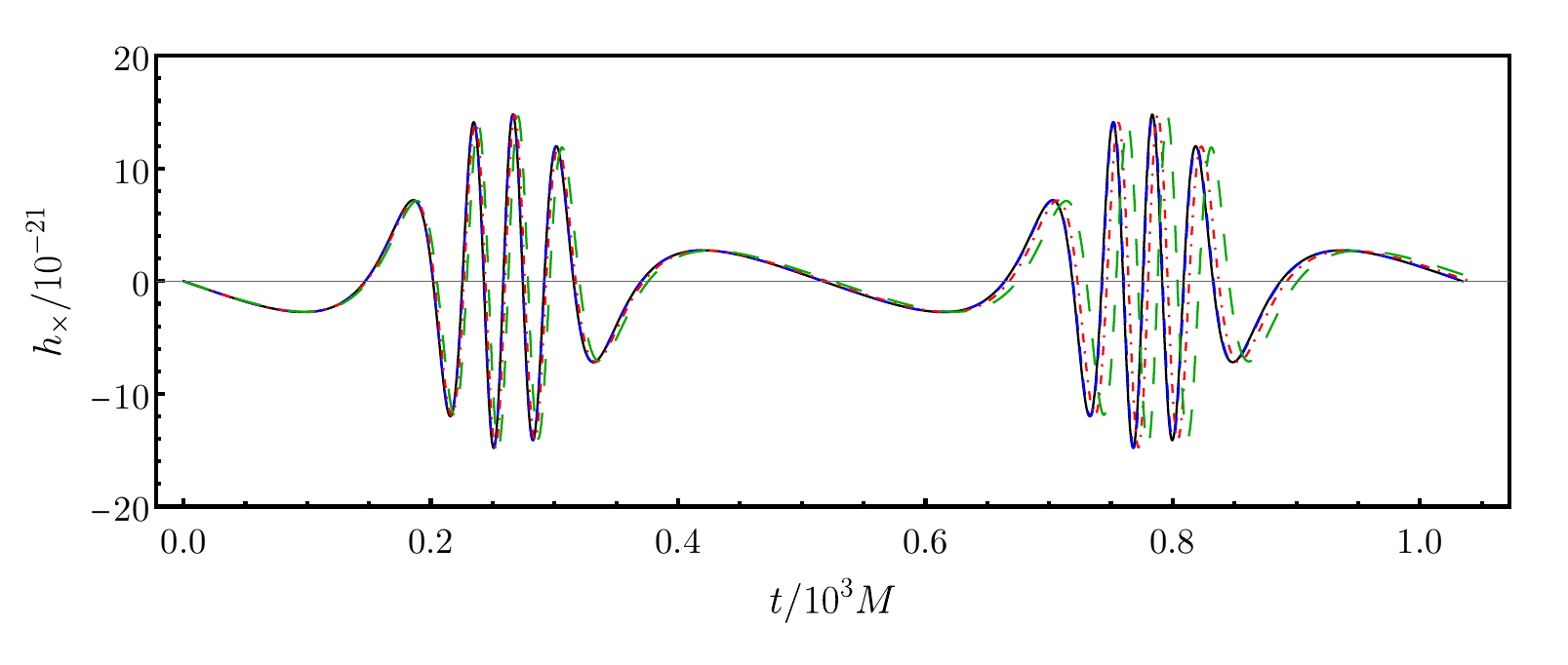}
   \includegraphics[width=150mm]{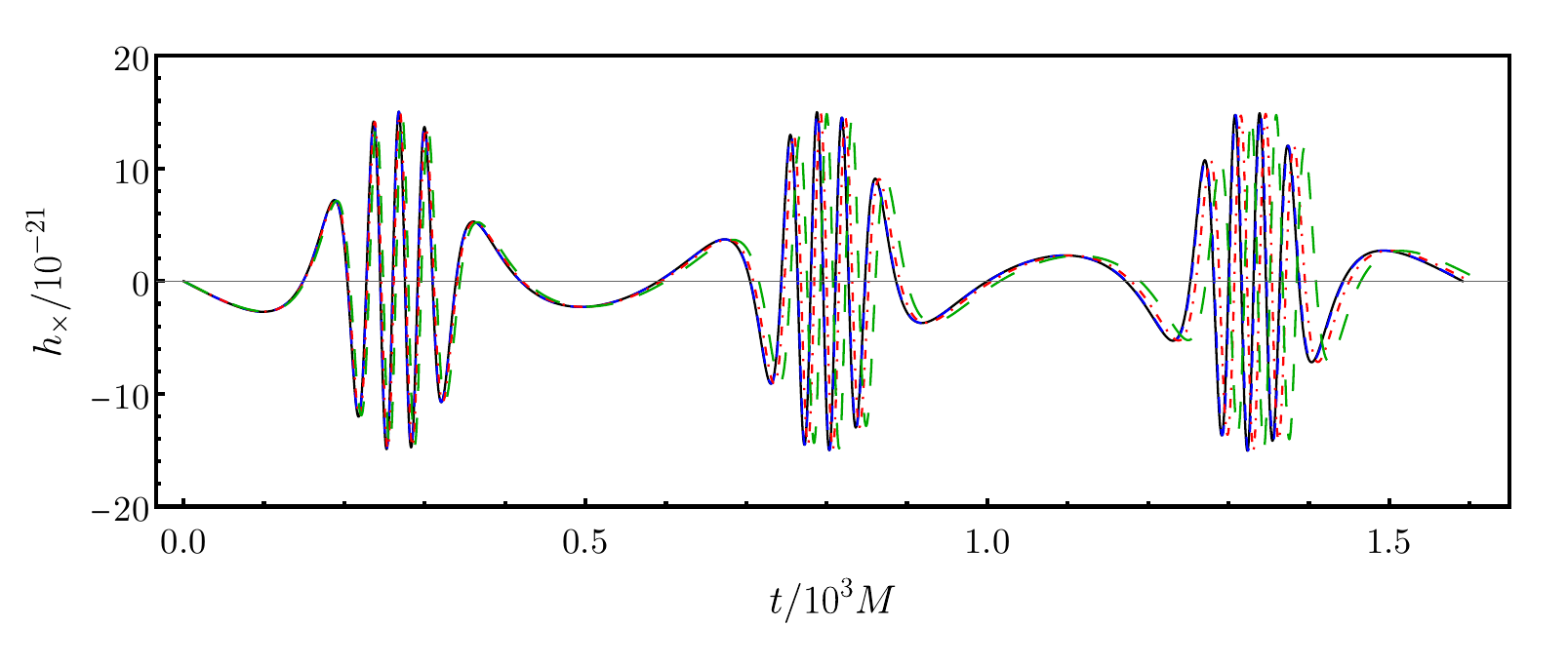}
   \caption{Cross polarization $h_\times$ for the periodic orbits $(1,1,0)$, $(2,1,1)$, and $(3,1,2)$. Line styles and parameter values follow Fig.~\ref{fig:comphplus}.}
    \label{fig:comphcross}
\end{figure}

Fig.~\ref{figs:colored} illustrates the relation between the orbital trajectory and its associated gravitational wave polarizations in the phantom DBI spacetime. The color-coded segments, corresponding to successive $\pi/2$ intervals of the orbital phase, allow the waveform features to be traced to different portions of the trajectory. The signals exhibit a characteristic modulation associated with the zoom--whirl motion, with stronger amplitudes during close passages and weaker emission as the particle moves toward apoapsis. Variations in the radial and azimuthal velocities further influence the oscillatory structure of both polarization modes. To explore the role of the regular core, we next compare the resulting periodic orbits and waveforms for different values of the regular core parameter $a$ for three representative periodic
orbits, $(1,1,0)$, $(2,1,1)$, and $(3,1,2)$ in Fig.~\ref{fig:comporbit}--\ref{fig:comphcross}.
As $a$ increases, the trajectories expand slightly,
while their characteristic zoom and whirl structures
remain largely preserved. The corresponding gravitational
wave polarizations, $h_+$ and $h_\times$, exhibit a
marginal decrease in amplitude, accompanied by a
slight increase in the oscillation period. However, the small
differences in their periods lead to a gradually
increasing phase offset over successive cycles.
These results suggest that, within the parameter
range considered, the regular core has a limited
effect on the overall orbital and waveform morphology,
with its influence becoming more apparent through
the accumulated phase difference than through
variations in the signal amplitude.

\section{Conclusions}
\label{sec:conclusion}

In this work, we have examined periodic orbits and their associated gravitational wave emission in the spacetime of a regular black hole supported by a phantom Dirac--Born--Infeld scalar field. Our results show that the regular core parameter $a$ introduces small but systematic changes in the orbital dynamics. For a given $(z,w,v)$ configuration, increasing $a$ leads to slightly higher orbital energies and more extended trajectories, without substantially altering their characteristic zoom--whirl patterns. The precession parameter grows with orbital energy, becoming increasingly sensitive as the separatrix is approached.

We also investigated how small energy perturbations affect periodic motion. Even minor deviations from the energy required for exact closure can noticeably alter the orbital pattern, particularly for trajectories with larger precession parameters.

The observed periastron advance of the S2 star around Sgr~A$^*$ was used to assess possible constraints on the regular core parameter. Although the DBI contribution produces an additional prograde precession, its magnitude remains below the current observational uncertainty throughout the black hole branch. The S2 measurements therefore provide no stronger restriction on $a$ than the condition for the existence of an event horizon.

Turning to gravitational radiation, we calculated the waveforms associated with representative periodic trajectories using a quadrupole-based numerical-kludge approach. As $a$ increases, the waveform amplitudes decrease slightly, while the oscillation periods become longer. These subtle changes give rise to a growing phase difference over successive cycles, indicating that the regular core leaves a more pronounced imprint on the waveform's phase evolution than on its amplitude.

Taken together, these results establish a connection between the regular core structure, strong--field orbital motion, and the resulting gravitational radiation. While the deviations from Schwarzschild remain modest in the parameter range explored, their cumulative effect on the waveform phase may offer a useful avenue for further investigation. A more precise assessment of their observational significance will require the inclusion of radiation reaction and more accurate relativistic waveform models.

\section*{Acknowledgments}
\hspace{0.5cm} 
N. H is supported by the Conselho Nacional de Desenvolvimento Científico e Tecnológico (CNPq), grant No. 152891/2025-0. 
N. H. also acknowledges the networking support provided by COST Action CA22113 -- Fundamental challenges in theoretical physics (Theory and Challenges), CA21106 -- COSMIC WISPers in the Dark Universe: Theory, astrophysics and experiments (CosmicWISPers), CA21136 -- Addressing observational tensions in cosmology with systematics and fundamental physics (CosmoVerse), and CA23130 -- Bridging high and low energies in search of quantum gravity (BridgeQG). A.A.A.F. is supported by Conselho Nacional de Desenvolvimento Cient\'{\i}fico e Tecnol\'{o}gico (CNPq) with the project number 150223/2025-0. I. P. L. was partially supported by the National Council for Scientific and Technological Development - CNPq, grant 312547/2023-4 and to acknowledge networking support by the COST Action BridgeQG (CA23130), the COST Action RQI (CA23115) and the COST Action FuSe (CA24101) supported by COST (European Cooperation in Science and Technology).  V. B. Bezerra is partially supported by Conselho Nacional de Desenvolvimento Cient\'{\i}fico e Tecnol\'{o}gico (CNPq) - Brazil, Grant number 311847/2026-9.


	\bibliography{main}

\begin{thebibliography}{10}

\bibitem{parvez2026exact}
Tausif Parvez and S~Shankaranarayanan.
\newblock Exact, nonsingular black holes from a phantom dbi field as primordial dark matter.
\newblock {\em Physical Review D}, 113(10):L101503, 2026.

\bibitem{ahmed2026hawking}
Faizuddin Ahmed, Ahmad Al-Badawi, and Edilberto~O Silva.
\newblock Hawking temperature, sparsity and energy emission rate of regular black holes supported by primordial dark matter.
\newblock {\em arXiv preprint arXiv:2605.04923}, 2026.

\bibitem{skvortsova2026massive}
Milena Skvortsova.
\newblock Massive scalar quasinormal modes of an asymptotically flat regular black hole supported by a phantom {Dirac--Born--Infeld} field.
\newblock {\em Ann. Phys.}, 492:170587, 2026.

\bibitem{lutfuouglu2026scalar}
Bekir~Can L{\"u}tf{\"u}o{\u{g}}lu.
\newblock Scalar, electromagnetic, and {Dirac} perturbations of regular black holes constituting primordial dark matter.
\newblock {\em J. Cosmol. Astropart. Phys.}, 2026(07):003, 2026.

\bibitem{abdullaev2026eikonal}
Mardon Abdullaev, Gumisbek Allambergenov, Diyorbek Rashidov, Pakhlavon Jamolov, and Javlon Rayimbaev.
\newblock Eikonal ringing, shadows, lensing, grey-body factors, and binding energy of asymptotically flat regular black holes in phantom {Dirac--Born--Infeld} gravity.
\newblock {\em Int. J. Gravit. Theor. Phys.}, 2(2):3, 2026.

\bibitem{heidari2026exploring}
N~Heidari, Iarley~P Lobo, VB~Bezerra, et~al.
\newblock Exploring a phantom dirac-born-infeld regular black hole via particle emission, wave scattering and geodesics.
\newblock {\em arXiv preprint arXiv:2608.23861}, 2026.

\bibitem{Addazi:2021xuf}
A.~Addazi et~al.
\newblock {Quantum gravity phenomenology at the dawn of the multi-messenger era{\textemdash}A review}.
\newblock {\em Prog. Part. Nucl. Phys.}, 125:103948, 2022.

\bibitem{Gong:2025mne}
Huajie Gong, Sheng Long, Xi-Jing Wang, Zhongwu Xia, Jian-Pin Wu, and Qiyuan Pan.
\newblock {Gravitational waveforms from periodic orbits around a novel regular black hole}.
\newblock 9 2025.

\bibitem{Kumar:2024our}
Shailesh Kumar and Tieguang Zi.
\newblock {Gravitational waves from regular black holes in extreme mass-ratio inspirals}.
\newblock {\em JCAP}, 10:108, 2025.

\bibitem{HEIDARI2026170773}
N.~Heidari, A.~A. Ara{\'u}jo~Filho, I.~P. Lobo, and V.~B. Bezerra.
\newblock Gravitational wave signatures from periodic orbits around a non–commutative schwarzschild black hole.
\newblock {\em Annals of Physics}, page 170773, 2026.

\bibitem{Ahmed:2025azu}
Fazlay Ahmed, Qiang Wu, Sushant~G. Ghosh, and Tao Zhu.
\newblock {Gravitational wave signatures from periodic orbits around a non-commutative inspired black hole surrounded by quintessence}.
\newblock {\em JCAP}, 02:004, 2026.

\bibitem{Chen:2025aqh}
Jiawei Chen and Jinsong Yang.
\newblock {Periodic orbits and gravitational waveforms in quantum-corrected black hole spacetimes}.
\newblock {\em Eur. Phys. J. C}, 85(7):726, 2025.

\bibitem{Ahmed:2025shr}
Fazlay Ahmed, Qiang Wu, Sushant~G. Ghosh, and Tao Zhu.
\newblock {Signatures of Quantum-Corrected Black Holes in Gravitational Waves from Periodic Orbits}.
\newblock 12 2025.

\bibitem{Alloqulov:2025dqi}
Mirzabek Alloqulov, Sanjar Shaymatov, Bobomurat Ahmedov, and Tao Zhu.
\newblock {Gravitational waveforms from periodic orbits around a dyonic ModMax black hole}.
\newblock {\em Eur. Phys. J. C}, 86(3):259, 2026.

\bibitem{Zhang:2025wni}
Chao Zhang and Tao Zhu.
\newblock {Periodic orbits and their gravitational wave radiations in $\gamma $-metric}.
\newblock {\em Eur. Phys. J. C}, 86(4):350, 2026.

\bibitem{Lu:2025cxx}
Shuo Lu and Tao Zhu.
\newblock {Gravitational radiations from periodic orbits around Einstein-{\AE}ther black holes}.
\newblock {\em Phys. Dark Univ.}, 50:102141, 2025.

\bibitem{Shi:2026zxx}
Zijian Shi, Xiangdong Zhang, and Yunlong Liu.
\newblock {Periodic orbits and gravitational waveforms of black holes in bumblebee gravity}.
\newblock 3 2026.

\bibitem{Liu:2025swi}
Mao-Jiang Liu, Long-Xing Huang, Yong-Qiang Wang, and Ke~Yang.
\newblock {Boson Stars in Bumblebee Gravity and Their Gravitational Waveforms from Extreme-Mass-Ratio Inspirals}.
\newblock 12 2025.

\bibitem{Junior:2024tmi}
Ednaldo L.~B. Junior, Jos{\'e} Tarciso S.~S. Junior, Francisco S.~N. Lobo, Manuel~E. Rodrigues, Diego Rubiera-Garcia, Lu{\'\i}s F.~Dias da~Silva, and Henrique~A. Vieira.
\newblock {Periodical orbits and waveforms with spontaneous Lorentz symmetry-breaking in Kalb{\textendash}Ramond gravity}.
\newblock {\em Eur. Phys. J. C}, 85(5):557, 2025.

\bibitem{Xia:2025yzg}
Zhong-Wu Xia, Huajie Gong, Qiyuan Pan, and Jiliang Jing.
\newblock {Probing Kalb{\textendash}Ramond field with extreme mass ratio inspirals}.
\newblock {\em Eur. Phys. J. C}, 85(9):960, 2025.

\bibitem{heidari2026gravitational}
N~Heidari, AA~Ara{\'u}jo~Filho, and Iarley~P Lobo.
\newblock Gravitational wave signatures and periodic orbits of a charged black hole in a hernquist dark matter halo.
\newblock {\em The European Physical Journal C}, 86(9):1104, 2026.

\bibitem{Deng:2025wzz}
Weike Deng, Sheng Long, Qin Tan, and Jiliang Jing.
\newblock {Gravitational waveforms from periodic orbits around a charged black hole with scalar hair*}.
\newblock {\em Chin. Phys.}, 50(3):035103, 2026.

\bibitem{Li:2025sfe}
Yong-Zhuang Li and Xiao-Mei Kuang.
\newblock {The bound orbits and gravitational waveforms of timelike particles around renormalization group improved Kerr black holes}.
\newblock {\em Eur. Phys. J. C}, 86(3):261, 2026.

\bibitem{kumar2024exploring}
Shailesh Kumar, Rishabh~Kumar Singh, Abhishek Chowdhuri, and Arpan Bhattacharyya.
\newblock Exploring waveforms with non-gr deviations for extreme mass-ratio inspirals.
\newblock {\em Journal of Cosmology and Astroparticle Physics}, 2024(10):047, 2024.

\bibitem{Wang:2025hla}
Chao-Hui Wang, Xiang-Cheng Meng, Yu-Peng Zhang, Tao Zhu, and Shao-Wen Wei.
\newblock {Equatorial periodic orbits and gravitational waveforms in a black hole free of Cauchy horizon}.
\newblock {\em JCAP}, 07:021, 2025.

\bibitem{Hua:2026kvw}
Zhutong Hua, Zhen-Tao He, Jing-Qi Lai, Jiageng Jiao, and Yu~Tian.
\newblock {Taxonomy of periodic orbits and gravitational waves in a non-rotating Destounis-Suvorov-Kokkotas black hole spacetime}.
\newblock {\em Phys. Lett. B}, 876:140402, 2026.

\bibitem{Xamidov:2026kqs}
Tursunali Xamidov, Sanjar Shaymatov, Qiang Wu, and Tao Zhu.
\newblock {Gravitational wave signatures from periodic orbits around a Schwarzschild-Bertotti-Robinson black hole}.
\newblock 2 2026.

\bibitem{levin2008periodic}
Janna Levin and Gabe Perez-Giz.
\newblock A periodic table for black hole orbits.
\newblock {\em Physical Review D}, 77(10):103005, 2008.

\bibitem{abuter2020detection}
R~Abuter, A~Amorim, M~Baub{\"o}ck, JP~Berger, H~Bonnet, W~Brandner, V~Cardoso, Y~Cl{\'e}net, PT~De~Zeeuw, J~Dexter, et~al.
\newblock Detection of the schwarzschild precession in the orbit of the star s2 near the galactic centre massive black hole.
\newblock {\em Astronomy \& Astrophysics}, 636:L5, 2020.

\bibitem{yao2026probing}
Jin-Tao Yao and Xin Li.
\newblock Probing covariant quantum black holes in effective quantum gravity through the orbital motion of the s2 star around sagittarius a$^*$.
\newblock {\em Physical Review D}, 113(8):084003, 2026.

\bibitem{li2024precessing}
Yong-Zhuang Li, Xiao-Mei Kuang, and Yu~Sang.
\newblock Precessing and periodic timelike orbits and their potential applications in einsteinian cubic gravity.
\newblock {\em The European Physical Journal C}, 84(5):529, 2024.

\bibitem{lin2022precessing}
Hou-Yu Lin and Xue-Mei Deng.
\newblock Precessing and periodic orbits around lee--wick black holes.
\newblock {\em The European Physical Journal Plus}, 137(2):176, 2022.

\bibitem{cardoso2022gravitational}
Vitor Cardoso, Kyriakos Destounis, Francisco Duque, Rodrigo~Panosso Macedo, and Andrea Maselli.
\newblock Gravitational waves from extreme-mass-ratio systems in astrophysical environments.
\newblock {\em Physical Review Letters}, 129(24):241103, 2022.

\bibitem{rahman2024probing}
Mostafizur Rahman, Shailesh Kumar, and Arpan Bhattacharyya.
\newblock Probing astrophysical environment with eccentric extreme mass-ratio inspirals.
\newblock {\em Journal of Cosmology and Astroparticle Physics}, 2024(01):035, 2024.

\bibitem{hughes2000evolution}
Scott~A Hughes.
\newblock Evolution of circular, nonequatorial orbits of kerr black holes due to gravitational-wave emission.
\newblock {\em Physical Review D}, 61(8):084004, 2000.

\bibitem{hughes2001evolution}
Scott~A Hughes.
\newblock Evolution of circular, nonequatorial orbits of kerr black holes due to gravitational-wave emission. ii. inspiral trajectories and gravitational waveforms.
\newblock {\em Physical Review D}, 64(6):064004, 2001.

\bibitem{babak2007kludge}
Stanislav Babak, Hua Fang, Jonathan~R Gair, Kostas Glampedakis, and Scott~A Hughes.
\newblock {Kludge} gravitational waveforms for a test-body orbiting a {Kerr} black hole.
\newblock {\em Physical Review D}, 75(2):024005, 2007.

\bibitem{zhao2025periodic}
Lai Zhao, Meirong Tang, and Zhaoyi Xu.
\newblock Periodic orbits and gravitational wave radiation in short hair black hole spacetimes for an extreme mass ratio system.
\newblock {\em The European Physical Journal C}, 85(1):36, 2025.

\bibitem{poisson2014gravity}
Eric Poisson and Clifford~M Will.
\newblock {\em Gravity: Newtonian, post-newtonian, relativistic}.
\newblock Cambridge university press, 2014.

\bibitem{yang2025gravitational}
Sen Yang, Yu-Peng Zhang, Tao Zhu, Li~Zhao, and Yu-Xiao Liu.
\newblock Gravitational waveforms from periodic orbits around a quantum-corrected black hole.
\newblock {\em Journal of Cosmology and Astroparticle Physics}, 2025(01):091, 2025.

\end{thebibliography}
	\bibliographystyle{unsrt}
	
\end{document}